\documentclass[11pt,a4paper]{article}
\usepackage{jcappub}

\usepackage{epsfig}
\usepackage{amssymb}

\newcommand{\alt}{\mbox{\;\raisebox{.3ex}
  {$<$}$\!\!\!\!\!$\raisebox{-.9ex}{$\sim$}\;}}
\newcommand{\agt}{\mbox{\;\raisebox{.3ex}
  {$>$}$\!\!\!\!\!$\raisebox{-.9ex}{$\sim$}}\;}

\newcommand{\be}{\begin{equation}}
\newcommand{\ee}{\end{equation}}
\newcommand{\bea}{\begin{eqnarray}}
\newcommand{\eea}{\end{eqnarray}}

\newcommand{\gagamma}{g_{a\gamma}}
\newcommand{\ckcs}{counts~keV$^{-1}$~cm$^{-2}$~s$^{-1}$}
\begin{document}

\newcommand{\zgz}{$^1$}
\newcommand{\usc}{$^2$}
\newcommand{\lbnl}{$^3$}
\newcommand{\cern}{$^4$}
\newcommand{\demok}{$^5$}
\newcommand{\irfu}{$^6$}
\newcommand{\tud}{$^7$}
\newcommand{\rbi}{$^8$}
\newcommand{\llnl}{$^9$}
\newcommand{\mpi}{$^{10}$}
\newcommand{\bgu}{$^{11}$}
\newcommand{\inr}{$^{12}$}
\newcommand{\nps}{$^{13}$}
\newcommand{\pat}{$^{14}$}




\title{
Towards a new generation \\axion helioscope}

\author{  I.~G.~Irastorza\zgz, F.~T.~Avignone\usc, S.~Caspi\lbnl, J.~M.~Carmona\zgz, T.~Dafni\zgz, M.~Davenport\cern, A.~Dudarev\cern, G.~Fanourakis\demok, E.~Ferrer-Ribas\irfu, J.~Gal\'an\zgz$^,$\irfu, J.~A.~Garc\'ia\zgz,
T.~Geralis\demok, I.~Giomataris\irfu, H.~G\'omez\zgz,
D.~H.~H.~Hoffmann\tud, F.~J.~Iguaz\irfu, K.~Jakov\v{c}i\'{c}\rbi,
   M.~Kr\v cmar\rbi, B.~Laki{\' c}\rbi, G.~Luz\'on\zgz, M.~Pivovaroff\llnl, T.~Papaevangelou\irfu,  G.~Raffelt\mpi, J.~Redondo\mpi, A.~Rodr\'iguez\zgz, S.~Russenschuck\cern, J.~Ruz\cern, I.~Shilon\cern$^,$\bgu,
   H.~Ten~Kate\cern, A.~Tom\'as\zgz, S.~Troitsky\inr,  K.~van~Bibber\nps, J.~A.~Villar\zgz, J.~Vogel\llnl, L.~Walckiers\cern, K.~Zioutas\pat}

\affiliation{
  \zgz Laboratorio de F\'{\i}sica Nuclear y Astropart\'{\i}culas, Universidad
  de Zaragoza, Zaragoza, Spain\\
  \usc Department of Physics and Astronomy, University of South Carolina, Columbia, SC, USA\\
  \lbnl Lawrence Berkeley National Laboratory, Berkeley, CA 94720, USA \\
  \cern CERN, Geneva, Switzerland\\
  \demok National Center for Scientific Research "Demokritos", Athens, Greece \\
  \irfu IRFU, Centre d'\'Etudes Nucl\'eaires de Saclay (CEA-Saclay), Gif-
  sur-Yvette, France\\
  \tud Technische Universit\"at Darmstadt, IKP, Darmstadt, Germany\\
  \rbi Rudjer Bo\v{s}kovi\'{c} Institute, Zagreb, Croatia\\
  \llnl Lawrence Livermore National Laboratory, Livermore, CA, USA\\
  \mpi Max-Planck-Institut f\"ur Physik, Munich, Germany \\
  \bgu Physics Department, Ben-Gurion University of the Negev, Beer-Sheva 84105, Israel \\
  \inr Institute for Nuclear Research (INR), Russian Academy of Sciences, Moscow, Russia\\
  \nps Naval Postgraduate School, Monterey, CA, USA \\
  \pat University of Patras, Patras, Greece\\
}

\emailAdd{Igor.Irastorza@cern.ch}

\abstract{
We study the feasibility of a new generation axion helioscope, the
most ambitious and promising detector of solar axions to date. We
show that large improvements in magnetic field volume, x-ray
focusing optics and detector backgrounds are possible beyond those
achieved in the CERN Axion Solar Telescope (CAST). For hadronic
models, a sensitivity to the axion-photon coupling of
$\gagamma\gtrsim {\rm few} \times 10^{-12}$~GeV$^{-1}$ is
conceivable, 1--1.5 orders of magnitude beyond the CAST sensitivity.
If axions also couple to electrons, the Sun produces a larger flux
for the same value of the Peccei-Quinn scale, allowing one to probe
a broader class of models. Except for the axion dark matter
searches, this experiment will be the most sensitive axion search
ever, reaching or surpassing the stringent bounds from SN1987A and
possibly testing the axion interpretation of anomalous white-dwarf
cooling that predicts $m_a$ of a few meV. Beyond axions, this new
instrument will probe entirely unexplored ranges of parameters for a
large variety of axion-like particles (ALPs) and other novel
excitations at the low-energy frontier of elementary particle
physics.
}

\maketitle
\flushbottom

\section{Introduction}                        \label{sec:introduction}

The Peccei-Quinn (PQ) mechanism of dynamical symmetry
restoration~\cite{Peccei:1977ur,Peccei:1977hh} stands out as the most
compelling solution of the strong CP problem, i.~e.~why this discrete
symmetry is apparently not violated by the non-trivial vacuum
structure of quantum chromodynamics (QCD). Central to the PQ mechanism
is the
axion~\cite{Weinberg:1977ma,Wilczek:1977pj,Peccei:2006as,Kim:2008hd},
the Nambu-Goldstone boson of a new spontaneously broken symmetry
U(1)$_{\rm PQ}$, with properties closely related to those of the
neutral pion. All axion properties are governed by a large energy
scale $f_a$, the axion decay constant, that is closely related to the
scale of symmetry breaking. The axion mass is given by $m_a f_a\sim
m_\pi f_\pi$, where $m_\pi=135$~MeV and $f_\pi=92$~MeV are the pion
mass and decay constant, respectively. The axion couplings with matter
and radiation also scale as $1/f_{\rm a}$. Experimental and
astrophysical constraints, if taken at face value, imply that $f_a\gtrsim10^9$~GeV,
corresponding to $m_a\lesssim10$~meV~\cite{Raffelt:2006cw}, and so
axions, despite their QCD origin, would be very light and very weakly
interacting, with interaction cross sections much smaller than those
of neutrinos. On the other hand, the unusual properties of axions
allow them to be produced in the early universe as coherent field
oscillations and as such to provide all or part of the cold dark
matter~\cite{Sikivie:2006ni,Wantz:2009it}.

It is still possible to find these ``invisible axions'' in
realistic search experiments and in this way test a fundamental
aspect of QCD. The generic $a\gamma\gamma$ vertex allows for
axion-photon conversion in external electric or magnetic fields in
analogy to the Primakoff effect for neutral pions. As shown in
1983 by Pierre Sikivie, the smallness of the axion mass allows
this conversion to take place coherently over macroscopic
distances, compensating for the smallness of the interaction
strength~\cite{Sikivie:1983ip}. Especially promising is to use the
Sun as a source for axions produced in its interior by the
Primakoff effect. Directing a strong dipole magnet toward the Sun
allows one to search for keV-range x-rays produced by axion-photon
conversion, a process best visualized as a particle oscillation
phenomenon~\cite{Raffelt:1987im} in analogy to neutrino flavor
oscillations. Three such helioscopes have been built, in
Brookhaven~\cite{Lazarus:1992ry}, Tokyo~\cite{Moriyama:1998kd,Inoue:2002qy,Inoue:2008zp} and
at CERN~\cite{Zioutas:1998cc}. The CERN Axion Solar Telescope
(CAST) is currently finishing a 8-year long data taking period,
having strongly improved on previous experiments and even
surpassed astrophysical limits in some range of parameters,
although axions have not been found.

One major difficulty with the helioscope technique is that for the
smallest axion-photon interaction strength $g_{a\gamma}\sim
\alpha/(2\pi f_a)\sim 10^{-10} $ GeV$^{-1}$ that has been accessible with CAST, the
corresponding axion mass $m_a\sim m_\pi f_\pi/f_a \sim $ eV is not small
enough to use the full coherent enhancement of the 10~m long magnet,
simply because the required axion-photon momentum transfer is too
large compared with the inverse length of the magnet. To overcome
this limitation, the conversion pipe was filled with a buffer gas,
providing the photons with a refractive mass $m_\gamma$ and
achieving the maximum conversion rate for a narrow $m_a$ range
around the $m_\gamma$ value defined by the gas pressure. CAST
has reached the ``axion line'' defined by $m_a f_a\sim m_\pi f_\pi$ in a
narrow range of masses below 1~eV by scanning many pressure
settings, but of course at the price of a reduced exposure time.

However, in this paper we show that large improvements in magnetic field volume,
x-ray focusing optics and detector backgrounds with respect to CAST
are possible. Based on these improvements, a new generation axion
helioscope (NGAH) could search for axions that are 1--1.5 orders of
magnitude more weakly interacting that those allowed by current CAST
constraints. If the ambitious goals defined in our study can be
achieved, a much larger range of realistic axion models can be
probed and it is even conceivable that one can reach a sensitivity
corresponding to $m_a$ in the 10~meV range. This mass range would be
significant in several ways. The energy-loss limit from SN~1987A
suggests that QCD axions have $f_a\gtrsim10^9$~GeV or
$m_a\lesssim10$--20~meV as mentioned earlier. Moreover, if axions also
interact with electrons, axions nearly saturating the SN~1987A limit
could explain the apparent anomalous energy loss of white
dwarfs~\cite{Isern:2010wz,Isern1992,Isern:2008nt,Isern:2008fs}. On
the experimental side, if the magnet length is 10~m as in CAST, the
sensitivity loss caused by the axion-photon momentum transfer begins
at $m_a\gtrsim20$~meV. In other words, if one can ``cross the axion
line'' at around this mass means that one can probe a large range of
axion models without buffer gas filling or with only few simple
pressure settings. For the first time, it appears conceivable to surpass the SN~1987A
constraint, test the white-dwarf cooling hypothesis, and begin to
explore entirely uncharted axion territory experimentally.

The tight connection of axions to QCD strongly restricts their
properties, leaving essentially only the one parameter $f_a$
undetermined, except of course for various model-dependent
numerical coefficients. On the other hand, the helioscope search
covers a much broader class of models where the mass and coupling
strength are independent parameters. Novel particles at the
low-energy frontier of high-energy physics are known under the
generic term Weakly Interacting Slim Particles (WISPs), axions
remaining a prime example. Extensions of the standard model often
include other very light pseudo-scalars and scalars coupled to two
photons that can be searched with helioscopes as well: majorons,
familons, dilatons, quintessence fields, and so forth. We call
these miscellanea ``axion-like particles" (ALPs). The current CAST
results have been used already to constrain further WISPs such as
weakly coupled hidden
photons~\cite{Redondo:2008aa,Gninenko:2008pz}. The proposed NGAH
could be extremely valuable to test further these scenarios and
survey the existence of other exotica such as
chameleons~\cite{Brax:2010xq}, mini-charged
particles~\cite{Holdom:1985ag,Davidson:1993sj} or more involved
ALP models which have been recently invoked to understand some
puzzling features of solar dynamics~\cite{Zioutas:2010tp,
Zioutas:2009bw, Zioutas:2007xk}. The motivations for various new
particles at the low-energy frontier of elementary particle
physics, their embedding and role in extensions of the standard
model, and our current knowledge about them have been reviewed
recently~\cite{Jaeckel:2010ni}. For the sake of simplicity, in
this paper we focus on axions and ALPs.

We begin our study in section~\ref{sec:hunting} with the current
status of axion and ALP searches, setting the goals for the NGAH.
In section~\ref{sec:helioscope} we describe the helioscope
technique and show the reach of a NGAH. In the following sections
we describe how to reach the required improvements in magnetic
field (section~\ref{sec:magnet}), x-ray optics
(section~\ref{optics}) and detectors (section~\ref{detectors}).
Our conclusions are presented in section~\ref{sec:conclusions}.

\section{Hunting Axions and ALPs}
\label{sec:hunting}

\subsection{Axion mass and interactions}

Most practical axion search strategies and all ALP searches are
based on the generic $a\gamma\gamma$ vertex which is usually written
as
\begin{equation}
{\cal L}_{a\gamma\gamma}=-\frac{C_\gamma \alpha}{8\pi f_a}F_{\mu\nu}\widetilde F^{\mu\nu}\; a
\equiv-\frac{\gagamma}{4}F_{\mu\nu}\widetilde F^{\mu\nu}\;a
=\gagamma\,{\bf E}\cdot{\bf B}\,a\,,
\end{equation}
where $F_{\mu\nu}$ and $\widetilde F_{\mu\nu}$ are the
electromagnetic field tensor and its dual and $a$ is the axion
field. $C_\gamma$ is a model dependent parameter given by
$C_\gamma\simeq E/N-2(4m_d+m_u)/3(m_u+m_d)\simeq E/N-1.92$ where
$E/N$ is the ratio of the electromagnetic and color anomalies of
the PQ symmetry, whereas $m_u$ and $m_d$ are the up and down quark
masses. One generic case is $E/N=0$, where the $a\gamma\gamma$
vertex derives exclusively from $a$-$\pi^0$-$\eta$ mixing, the
KSVZ model being a classic
example~\cite{Kim:1979if,Shifman:1979if}. Non-vanishing $E/N$
values derive from triangle-loop diagrams involving ordinary or
exotic particles carrying Peccei-Quinn charges. One generic case
is $E/N=8/3$ or $C_\gamma\simeq0.75$, relevant for models in grand
unified theories (GUTs), with the DFSZ model being the usual
example~\cite{Dine:1981rt,Zhitnitsky:1980tq}.  For general ALPs,
$\gagamma$ is the central parameter and not directly interpreted
in terms of some underlying~$f_a$.

For axions, in contrast to general ALPs, $\gagamma$ and $m_a$ are
closely related, both deriving from the same $a$-$\pi^0$-$\eta$
mixing at the core of the PQ mechanism. One finds
\begin{equation}
m_a=\frac{m_u+m_d}{\sqrt{m_u m_d}}\,\frac{m_\pi f_\pi}{f_a}
=6~{\rm meV}\,\frac{10^9~{\rm GeV}}{f_a}\,.
\end{equation}
For the numerical estimate we have used the canonical value for the
quark mass ratio $z=m_u/m_d=0.56$. The allowed range $z=0.35$--0.60
\cite{Nakamura:2010zzi} leads to about a 10\% uncertainty that we will
henceforth ignore.

The role of axions in QCD implies that they must interact with
hadrons and photons based on their generic $a$-$\pi^0$-$\eta$
mixing, even though one can construct models where some of these
couplings can be accidentally small by cancelation effects. In
addition, notably in GUT models, axions can interact with leptons.
If such interactions are absent, we speak of hadronic axions, if
lepton couplings exist, of non-hadronic axions. For the latter, the
axion-electron interaction is of greatest practical interest.
Usually it is written in the axial-vector derivative form
\begin{equation}
{\cal L}_{aee}=C_e\frac{\partial_\mu a}{2 f_a}\,\bar \psi_e \gamma_5\gamma^\mu \psi_e
\end{equation}
where $C_e$ is a model-dependent numerical coefficient. In many
situations the derivative interaction is equivalent to pseudoscalar
structure $g_{ae}\bar \psi_e \gamma_5 \psi_e\,a$, motivating us to
define the dimensionless axion-electron Yukawa coupling
$g_{ae}=C_e m_e/f_a$ with $m_e$ being the electron mass. In the
DFSZ model, we have $C_e=\frac{1}{3}\cos^2\beta$
where the denominator 3 arises from three particle families and
$\tan\beta$ is the ratio of expectation values of two Higgs fields
giving masses to up-type and down-type fermions, respectively.

Notice that hadronic axions, where $C_e=0$ at tree level, develop
radiatively a non-zero coupling to electrons, $C_e\propto C_\gamma
\alpha^2/2\pi$~\cite{Srednicki:1985xd}. This coupling does not seem to
be of practical interest for our study. In particular, axion
production in stars is always dominated by the tree-level couplings.

\subsection{Search strategies}

Axions are produced in the early universe by the misalignment
mechanism~\cite{Sikivie:2006ni,Wantz:2009it}. If the PQ symmetry is
restored by reheating after inflation, axion strings and domain walls
form and decay, providing an additional source of
axions. These relic axions provide the
cold dark matter for $m_a\sim 10~\mu$eV, with a large uncertainty in either direction, and actually $m_a$ could be as large as 200 $\mu$eV~\cite{Wantz:2009it}. In addition to
this ``classic axion window,'' the neV mass range or below has been
considered as the ``anthropic axion window''
\cite{Linde:1987bx,Hertzberg:2008wr} and a possible search strategy
was recently proposed \cite{Graham:2011qk}. In the classic window,
Sikivie's haloscope technique~\cite{Sikivie:1983ip}, based on axion
conversion into microwave photons, is the basis for the ongoing
large-scale Axion Dark Matter Experiment (ADMX)
\cite{Asztalos:2001tf,Asztalos:2003px}. If axions in the classic
window are the dark matter, they will be found.

In addition, axions with eV-range masses would have been produced
by thermal processes. They can contribute to the hot component of
the dark matter, similar to neutrinos~\cite{Moroi:1998qs}.
Hot-dark matter bounds based on precision cosmology require
$m_a\lesssim1$~eV \cite{Hannestad:2010yi}. This and other
complementary cosmological arguments allow one to exclude axions
in the entire $m_a$ range 1~eV--300~keV based on cosmological
evidence alone \cite{Cadamuro:2010cz}.

Axions or ALPs can also be produced and detected in the laboratory,
without invoking cosmological or astrophysical sources. Beam dump
experiments and rare decays have been used to rule out axions with
$m_a\gtrsim10$~keV. For smaller masses, the correspondingly smaller
couplings prevent one from studying invisible axion models with
traditional particle physics methods. For general ALPs where the
photon interaction strength is not related to the mass, the most
ambitious current initiatives involve photon regeneration
experiments (``light shining through a wall''), which are foreseen
to reach the $\gagamma\sim 10^{-11}$~GeV$^{-1}$ ballpark for
$m_a\lesssim1$~meV, but are unlikely to become sensitive to realistic
axion models (see Ref.~\cite{Redondo:2010dp} for a recent review).

\begin{figure}[t]
\centering
\includegraphics[width=8cm]{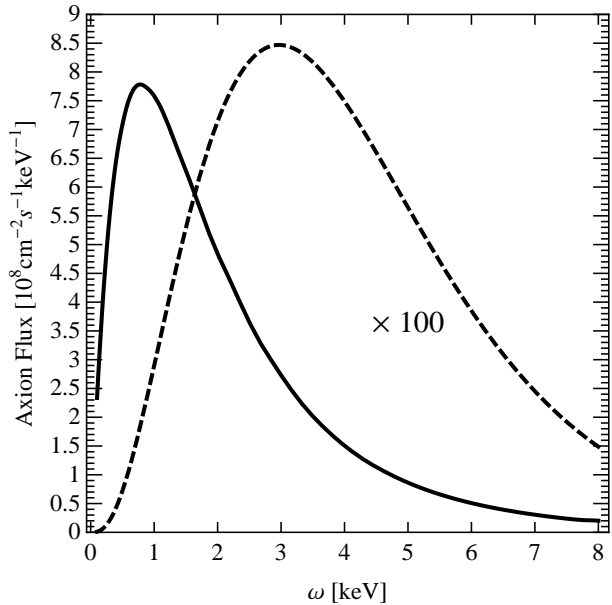}
\caption{\label{axion_flux} Solar axion flux spectrum at Earth,
originating from the Primakoff process (dashed line) and from
processes involving electrons (solid line), bremsstrahlung and
Compton processes. We have chosen illustrative values of
$g_{ae}=10^{-13}$ and $\gagamma=10^{-12}~{\rm GeV}^{-1}$,
corresponding to DFSZ axions with $f_a=0.85\times 10^9$~GeV,
$C_e=1/6$ and $C_\gamma=0.75$. For better comparison, the
Primakoff flux has been scaled up by a factor of 100.}
\end{figure}

Finally, axions are copiously produced in stellar interiors. By far
the brightest object in the sky is the Sun, and the same is true in
the light of neutrinos or axions, except of course at much higher
energies where non-thermal sources dominate. Axions can be produced
in the Sun through a variety of reactions. Hadronic axions and ALPs
are mainly produced via Primakoff conversion of thermal photons in
the Coulomb fields of charged particles via the $a\gamma\gamma$
vertex. The usual solar Primakoff spectrum peaks near the mean
energy of 4.2~keV and exponentially decreases at larger energies as
shown in figure~\ref{axion_flux}.

For non-hadronic axions, defined as having tree-level interactions
with electrons, the dominant emission processes are electron-nucleus
bremsstrahlung $e+Ze\to Ze+e+a$, electron-electron bremsstrahlung
$e+e\to e+e+a$, and the Compton process $\gamma+e\to e+a$. In
addition, free-bound transitions play a sub-dominant role. The
relative importance for solar energy loss of the first three reactions
is roughly 2:1:1, respectively~\cite{Raffelt:1985nk}. The resulting
spectrum is softer than the Primakoff one, with a mean energy of
1.8~keV and peaking below 1~keV~\cite{Raffelt:1986tq} as shown in
figure~\ref{axion_flux}. The integrated solar axion flux from the
electron coupling is much larger than from photon coupling
\begin{equation}
\label{ratio}
\frac{\Phi_{ae}}{\Phi_{a\gamma}}\sim
900 \left(\frac{C_e}{C_\gamma}\right)^2 \ .
\end{equation}
However, previous solar axion searches have relied primarily on the
Primakoff process in order to cover the broader class of ALPs. For
non-hadronic axions, astrophysical limits on the $g_{ae}$ from
globular cluster stars and white dwarfs are so restrictive that
detecting them from the Sun seemed hopeless. On the other hand, the
CAST limit on $\gagamma$ for low-mass ALPs actually supersedes
astrophysical limits from globular cluster stars. In the proposed
new helioscope, for the first time it becomes conceivable to
supersede even astrophysical limits on $g_{ae}$ and to probe new axion
territory for the broadest class of models.

By means of the axion-photon coupling, solar axions can be efficiently
converted back into photons in the presence of an electromagnetic
field. The energy of the reconverted photon is equal to the incoming
axion, so a flux of detectable x-rays of few keV energies is
expected. Crystal detectors may provide such
fields~\cite{Buchmuller:1989rb,Paschos:1993yf,Creswick:1997pg}, giving
rise to characteristic Bragg patterns that have been searched for
as byproducts of dark matter
searches~\cite{Avignone:1997th,Morales:2001we,Bernabei:2001ny,Collaboration:2009ht}.
However, the prospects of this technique have proven
limited~\cite{Cebrian:1998mu,Avignone:2010zn} and do not compete with
dedicated helioscope experiments.

\begin{figure}[t]
\centering
\includegraphics[width=14cm]{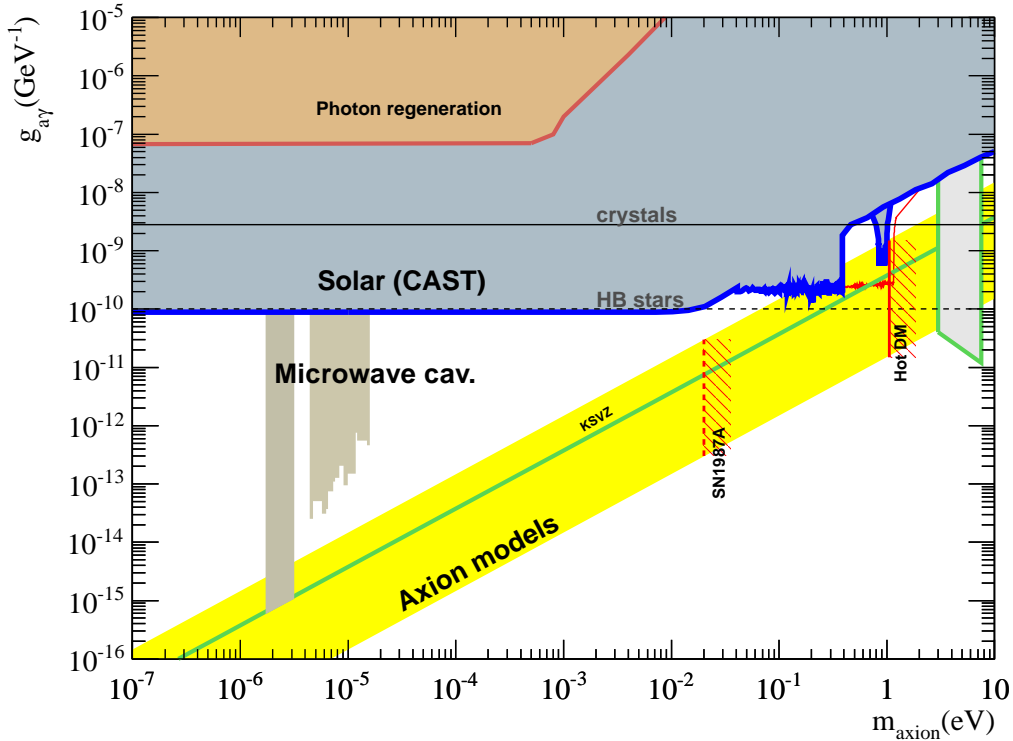}
\caption{\label{exclusion_large}Comprehensive ALP parameter space,
highlighting the three main front lines of direct detection
experiments: laser-based laboratory techniques, helioscope (solar
ALPs and axions), and microwave cavities (dark matter axions). The blue line corresponds to the current helioscope limits, dominated by CAST~\cite{Andriamonje:2007ew,Arik:2008mq} for practically all axion masses but for the $m_a \sim 0.85-1$ eV exclusion line from the last Tokyo helioscope results\cite{Inoue:2008zp}. Also shown are the constraints from horizontal branch (HB)
  stars, supernova SN1987A, and hot dark matter (HDM). The yellow ``axion band'' is defined roughly by $m_a
f_a\sim m_\pi f_\pi$ with a somewhat arbitrary width representing
the range of realistic models. The green line refers to the KSVZ
model ($C_\gamma\sim-1.92$).
}
\end{figure}

\subsection{Hadronic axions and ALPs}

The most relevant coupling of hadronic axions and the defining
property of ALPs is their two photon coupling. It is therefore natural
to discuss the status of ALP and hadronic axion searches in the
two-dimensional $m_a$-$\gagamma$ parameter space
(figure~\ref{exclusion_large}).  In this way we can clearly show three
main frontlines in the direct search for hadronic axions: dark matter
axions, solar axions and laboratory axions. Dark matter axion
experiments, of which ADMX is the only active example, are sensitive
down to very low $\gagamma$ values at the very low mass range broadly
circumscribed by 1--$100~\mu$eV. ADMX contemplates the exploration of
the 1--10~$\mu$eV decade with sufficient sensitivity in $\gagamma$ to
exclude or detect the QCD axion band, corresponding to $\gagamma$ in
the approximate range $10^{-17}$--$10^{-14}$~GeV$^{-1}$. And
subsequently, assuming the success of a dedicated R\&D program, the
technique could also be applied to the next decade in mass, up to
$100~\mu$eV. However, for axion masses above this value, or much below
$1~\mu$eV, the resonant cavity technique becomes impractical.

Solar axion searches, although not reaching such low $\gagamma$
values, are sensitive to a very wide range of axion and ALP
masses.  As detailed later on, the most recent CAST results set the
most stringent limits for ALP masses up to the eV scale.

On the heaviest end of this mass range, the situation is very
different for axions and ALPs. Hadronic axions are constrained by the
neutrino burst duration of SN 1987A to be lighter than 10--20~meV,
whereas ALPs are not subject to this bound~\cite{Masso:1995tw}. Of course, it may be
unwise to rely entirely on a single observation with sparse statistics
and intrinsic uncertainties~\cite{Raffelt:2006cw} to consider an
entire class of models excluded. Therefore, it is both of conceptual
and practical interest that in phase II of
CAST one begins to probe realistic axion models, even at the price of
the very cumbersome scanning through many pressure settings of the
buffer gas.

Pure laboratory searches are of great fundamental interest because
they do not depend on astrophysical or cosmological axion sources.
However, again because of the required axion-photon momentum transfer,
these optical-light experiments only reach to sub-meV masses and are
far away from competing with solar axion searches as seen in
figure~\ref{exclusion_large}.

Without any other realistic ideas on the market, pushing the perimeter
of ALP and hadronic axion sensitivity in the $m_a$-$g_{a\gamma}$
parameter space seems a task that only a new generation axion helioscope
can attain. We will
argue in the following sections that in the sub-eV mass range one will
be able to access $\gagamma$ sensitivities well below the
$10^{-11}$~GeV$^{-1}$ level and even approaching
$10^{-12}$~GeV$^{-1}$. This region covers an important fraction of the
remaining axion parameter space, not firmly excluded by astrophysical
considerations, and complementary to the region to be explored by
ADMX.

When it comes to ALPs, the region just beyond the current CAST
sensitivity has a special phenomenological interest.  Very small
mass ALPs with two-photon couplings in the $\gagamma$ ballpark of
$10^{-12}$--$10^{-10}$~GeV$^{-1}$ have been invoked in the context
of a number of puzzling astrophysical observations. Photons
propagating in galactic or intergalactic magnetic fields can
oscillate into ALPs -or vice versa-, altering the properties of
light propagation in our universe. Particularly interesting issues
are the observation of $\gamma$-rays,
e.~g.~\cite{Teshima:2007zw,Aharonian:2005gh}, and ultra high
energy cosmic rays~\cite{Gorbunov:2004bs,Abbasi:2005qy} from very
distant sources, such as active galactic nuclei.  Photon-ALP
conversion has been invoked by a number of
authors~\cite{Csaki:2003ef,DeAngelis:2008sk,Roncadelli:2008zz,Simet:2007sa,Fairbairn:2009zi,Albuquerque:2010rq}
to account for these observations via a photon regeneration
effect.

The photon-ALP mixing in the galactic or intergalactic
medium has other testable consequences. The random character of
astrophysical magnetic fields produces a particular scattering of the
photon arrival probability that can also be used to test the ALP
hypothesis~\cite{Mirizzi:2009aj}. Recently, some luminosity relations
of active galactic nuclei were shown to have precisely this particular
scatter~\cite{Burrage:2009mj} although this claim is still
controversial~\cite{Pettinari:2010ay}.  Finally, photon-ALP mixing is
polarization dependent, a fact that could explain long-distant
correlations of quasar polarization~\cite{Payez:2008pm} and offers
further testing opportunities~\cite{Bassan:2010ya}.

\subsection{Non-hadronic axions}

Axions with a tree level coupling to electrons have a somewhat
different physics case and phenomenology.  From the theoretical point
of view, these models are very appealing since they arise for instance
in grand unified theories (GUTs), strongly motivated completions of
the standard model at high energies.  From the phenomenological side
it is worth repeating that naturally the coupling to
electrons leads to larger axion fluxes from stars than the coupling to
photons, offering enhanced fluxes to a NGAH. Of course, this also
strengthens the purely astrophysical bounds.  In the case of the Sun,
agreement between the measured and predicted solar neutrino flux
constrains the energy loss in axions to be below 10\% of the solar
luminosity~\cite{Gondolo:2008dd}, which translates into a bound
$g_{ae}<1.4\times 10^{-11}$~\cite{Raffelt:1985nk}.  A more stringent
bound comes from the delay of helium ignition in red giants of
globular clusters, namely $g_{ae}<2.5\times
10^{-13}$~\cite{Raffelt:1989xu,Raffelt:1994ry}.

Axion emission is also constrained by the cooling of white dwarfs
(WDs)~\cite{Raffelt:1985nj}.  However, in this case the observed
WD luminosity function~\cite{Isern:2008nt,Isern:2008fs} seems to actually \emph{prefer} some axion emission. Independent evidence for an extra cooling mechanism is provided by the rate of change of the pulsation period of ZZ~Ceti star G117-B15A~\cite{Isern:2010wz,Isern1992}. Both
observations suggest values in the range $g_{ae}= 0.7$--$1.3\times 10^{-13}$.
Testing this hypothesis by means of a laboratory experiment is
extremely well motivated. Our proposed helioscope is probably the only
viable option for such a test.

In order to quantify the sensitivity of helioscopes to non-hadronic
axions we must consider the three parameters, $m_a$, $g_{a\gamma}$ and
$g_{ae}$.  However, the helioscope signal depends upon the combination
$(g_{ae}\gagamma)^2$, since the photon coupling determines the
detection rate and the electron coupling the axion production in the
Sun, unless the electron coupling is unnaturally small. We here do not
worry about the fine-tuned case where the photon and electron couplings
provide comparable solar fluxes.  
The sensitivity of helioscopes to non-hadronic
axions is then better appreciated in the $m_a$-$C_eC_\gamma$ plane shown
on the right panel of figure~
\ref{fig:scenarios}.
In order to compare with the
red-giant bound and the WD motivated region one must fix the parameter
$C_\gamma$. For non-hadronic models the GUT assumption $E/N=8/3$ is
probably the best motivated case, leading to $C_\gamma=0.75$ as
discussed earlier. With this assumption, the red giant bound is shown
as a diagonal dashed line in the plot on the right panel of figure~
\ref{fig:scenarios}, whereas the orange band of the same plot refers to the generous range $0.5\times10^{-13}<g_{ae}<2\times10^{-13}$, motivated by WD cooling. The DFSZ
models~\cite{Dine:1981rt,Zhitnitsky:1980tq}, for which $C_e=(\cos^2\beta)/3<1/3$,
are bound by the labeled horizontal line.
Although CAST sensitivity is not well tuned to search for
non-hadronic axions, we will show that advances surveyed in this paper could make a NGAH surpass the red giant constraints and start probing the region of parameter space highlighted by the cooling of WDs.


In summary, there is strong motivation to improve the helioscope
sensitivity beyond CAST to $\gagamma$ down to 10$^{-12}$~GeV$^{-1}$
and the $g_{ae}$ sensitivity down to $10^{-13}$. This region includes,
on the high-mass side, a large set of plausible QCD axion models, potentially superseding
the SN 1987A and red-giant bound in non-hadronic models and start
probing the parameters suggested by the longstanding anomalous WD
cooling.  At lower axion masses, this region includes the ALP
parameters invoked repeatedly by several works as being behind some
astrophysical phenomena. This region is currently not excluded by any
experimental results or astrophysical limits and moreover, it is
out of reach for other foreseeable experimental techniques, except maybe a
next generation~\cite{Mueller:2009wt,Arias:2010bh,Redondo:2010dp} of photon regeneration experiment using resonant techniques,
and this only in the sub meV mass range.
We will argue that for a new generation axion helioscope
such enhancements are
technically feasible, by reasonably extending the innovations
introduced by CAST.  We propose such an experiment as the
next large scale project that the experimental axion community should
envisage for the next decade.

\section{An enhanced axion helioscope}
\label{sec:helioscope}

The probability that an axion going through the transverse magnetic field $B$ over a
length $L$ will convert to a photon is given by \cite{Sikivie:1983ip,Zioutas:2004hi,Andriamonje:2007ew}:

\begin{eqnarray}\label{conversion_prob}
  P_{a\gamma} = 2.6 \times 10^{-17} \left(\frac{B}{10 \mathrm{\ T}}\right)^2
  \left(\frac{L}{10 \mathrm{\ m}}\right)^2 \nonumber \left(\gagamma \times 10^{10}
  \mathrm{\ GeV}\right)^2\mathcal{F}
\end{eqnarray}

\noindent where the form factor $\mathcal{F}$
accounts for the coherence of the process:

\begin{equation}\label{matrix_element}
    \mathcal{F}=\frac{2(1-\cos q L)}{(qL)^2}
\end{equation}

\noindent and $q$ is the momentum transfer. The fact that the axion
is not massless, puts the axion and photon waves out of phase
after a certain length. The coherence is preserved
($\mathcal{F} \simeq 1$) as long as $qL \ll 1$,
which for solar axion energies and a magnet length of 10 m
(like the one of CAST) happens at axion masses up to $\sim
10^{-2}$ eV, while for higher masses $\mathcal{F}$
begins to decrease, and so does the sensitivity of the experiment.
To mitigate the loss of coherence, a buffer gas can be introduced into the magnet
beam pipes \cite{Arik:2008mq} to impart an effective
mass to the photons $m_\gamma = \omega_p$
(where $\omega_p$ is the plasma frequency of the gas,
 $\omega_{\rm p}^2=4\pi\alpha n_e/m_e$).
For axion masses that match the photon mass, $q=0$ and the
coherence is restored. By changing the pressure of the gas inside
the pipe in a controlled manner, the photon mass can be
systematically increased and the sensitivity of the experiment can
be extended to higher axion masses.

The basic layout of an axion helioscope requires a powerful magnet
coupled to one or more x-ray detectors. When the magnet is aligned
with the Sun, an excess of x-rays at the exit of the magnet is
expected, over the background measured at non-alignment periods.
This detection concept was first experimentally implemented
in~\cite{Lazarus:1992ry} and later by the Tokyo
helioscope~\cite{Moriyama:1998kd}, which provided the first limit
to solar axions which is self-consistent, i.~e.~compatible with
solar physics. During the last decade, the same basic concept has
been used by CAST
\cite{Zioutas:1998cc,Zioutas:2004hi,Andriamonje:2007ew,Arik:2008mq}
with some innovations that provide a considerable step forward in
sensitivity to solar axions.

The CAST experiment is the most powerful axion helioscope ever constructed.
As the conversion magnet is the main driver of a helioscope's
sensitivity, the CAST collaboration has harnessed the most advanced superconducting
magnet technology of CERN. Specifically, CAST uses a decommissioned
LHC test magnet that provides a magnetic field of 9 Tesla along
its two parallel pipes of 2$\times$14.5 cm$^2$ area and 10 m
length, increasing the corresponding axion-photon conversion
probability by a factor 100 with respect to the previous
implementation of the helioscope concept \cite{Zioutas:2004hi}.
The magnet is able to track the Sun for $\sim$ 3 hours per day, half
in the morning at sunrise and half in the evening at sunset. The rest of the day is
used for background measurements. X-ray detectors
are placed at the ends of the bores, with a Micromegas detector
\cite{Abbon:2007ug} and a CCD \cite{Kuster:2007ue} installed at the ``sunrise'' side,
and two additional Micromegas detectors installed at the ``sunset'' side.
(In 2007, the sunset Mircomegas detectors replaced the
multiwire TPC \cite{Autiero:2007uf} that had been used previously.)

The unsurpassed sensitivity of CAST relies, in part, on several
pioneering enhancements to the helioscope concept. First,
CAST employs an x-ray focusing optic between the magnet and the detector,
focusing the putative x-ray signal to a small spot and thus increasing
the signal-to-background and sensitivity of the experiment. Additionally,
in the event of a positive signal and actual detection, such an optic would
become a real ``axion-imaging" telescope.
The CAST CCD is coupled to one such device~\cite{Kuster:2007ue},
a Wolter telescope borrowed from the field of x-ray astronomy, that enhances its
signal-to-background ratio by two orders of magnitude. Second,
CAST has actively and continually applied state-of-the-art low
background techniques to all its detector subsystems, in order to
minimize the experimental background and further increase the
sensitivity. These include the use of low radioactivity materials
for the detector components and surroundings, the simulation and
modeling of backgrounds, the use of shielding against external
radiation, and the development of sophisticated offline analysis
criteria to discriminate signal events and reject background.

The experiment released its initial results (phase I) from data
taken in 2003 and 2004 without buffer
gas~\cite{Zioutas:2004hi,Andriamonje:2007ew}. No signal above
background was observed. For hadronic axions this implies an upper
limit to the axion-photon coupling $\gagamma < 8.8 \times
10^{-11}~{\rm GeV}^{-1}$ at 95\% CL for the low mass (coherence)
region $m_a \alt 0.02~{\rm eV}$ (figure~\ref{exclusion_large}). In
2006, the experiment embarked on phase II operations, which
requires a buffer gas inside the magnet bores to recover the
coherence of the conversion for specific axion masses matching the
effective photon mass defined by the buffer gas density. The
pressure of the gas is changed in discrete small steps to scan the
parameter space above $m_a \sim 0.02$ eV. The data acquired in
2006 \cite{Arik:2008mq}, with $^4$He as the buffer gas, scanned
for axion masses up to 0.39 eV at a level for axion-photon
couplings down to $\sim$ $2.2 \times 10^{-10}~{\rm GeV}^{-1}$,
entering into the QCD axion model band, as shown in
figure~\ref{exclusion_large}. To gain access to higher masses, in
2007 the buffer gas was switched to $^3$He to avoid gas
condensation at the required pressure.  The experiment is
currently engaged in a systematic scan of axion masses above 0.4
eV. The $^3$He data taking began in 2008, and upon completion of
this research program in the middle of 2011, CAST will have
explored an axion mass up to $\sim$ 1.2 eV, overlapping with the
cosmological upper limit on the axion mass discussed above.


In summary, CAST has provided the best experimental limit on
$\gagamma$ for a wide range of axion masses, up to $\sim$ 0.02 eV,
a result that now supersedes the astrophysical limit derived from
energy-loss arguments on globular cluster stars. In its second
phase, CAST has been configured to be sensitive to hadronic axion
models, in the mass region just below 1 eV. During its last years
of operation, CAST has not only built the largest and most
sensitive axion helioscope, but has also improved on the original
concept of an axion helioscope  and developed the expertise that
will be crucial for a marked gain in sensitivity, as envisioned
with a NGAH.


\subsection{Figures of merit}

\begin{figure}[t] \centering
\includegraphics[width=5cm]{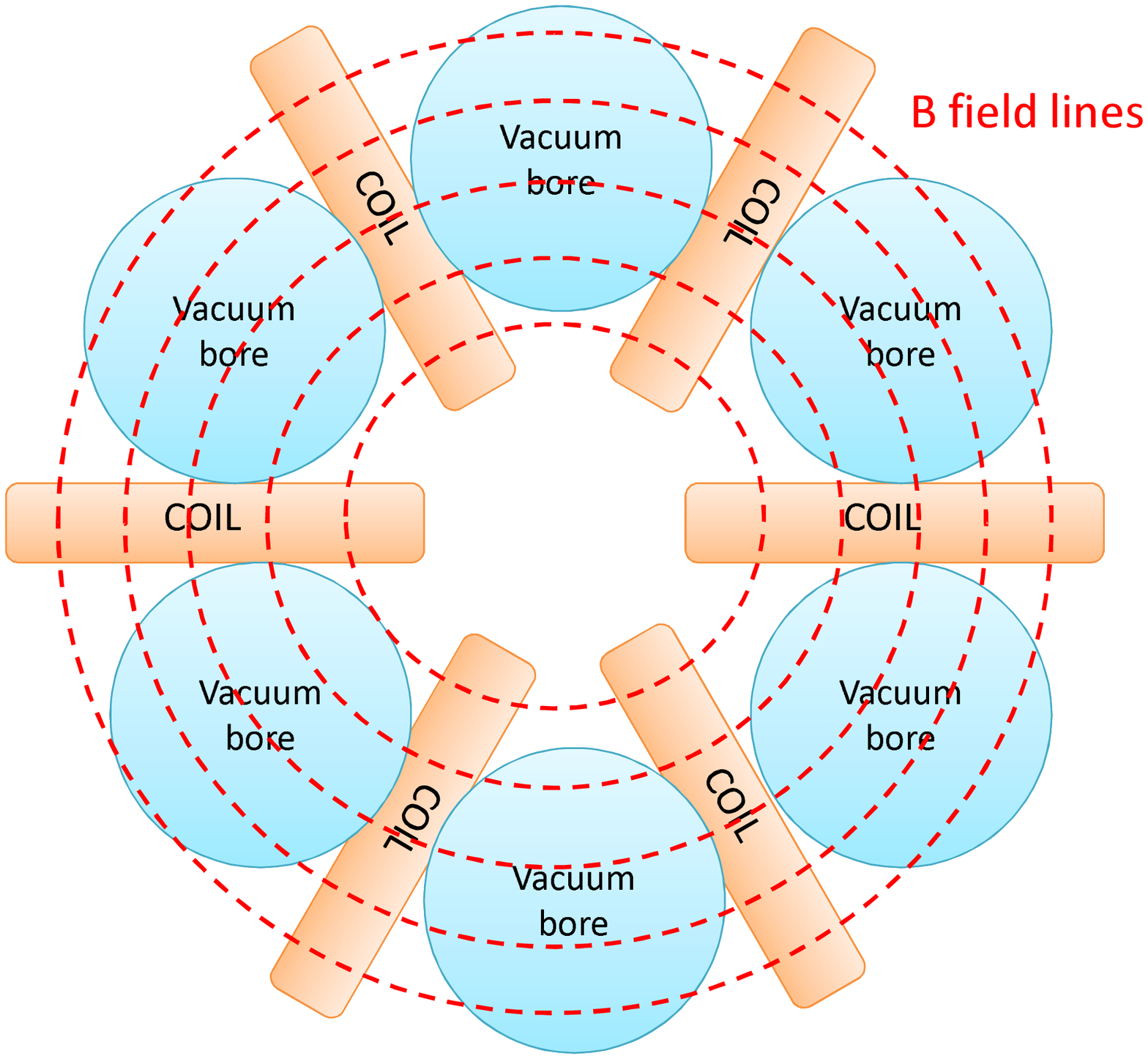}
\includegraphics[width=10cm]{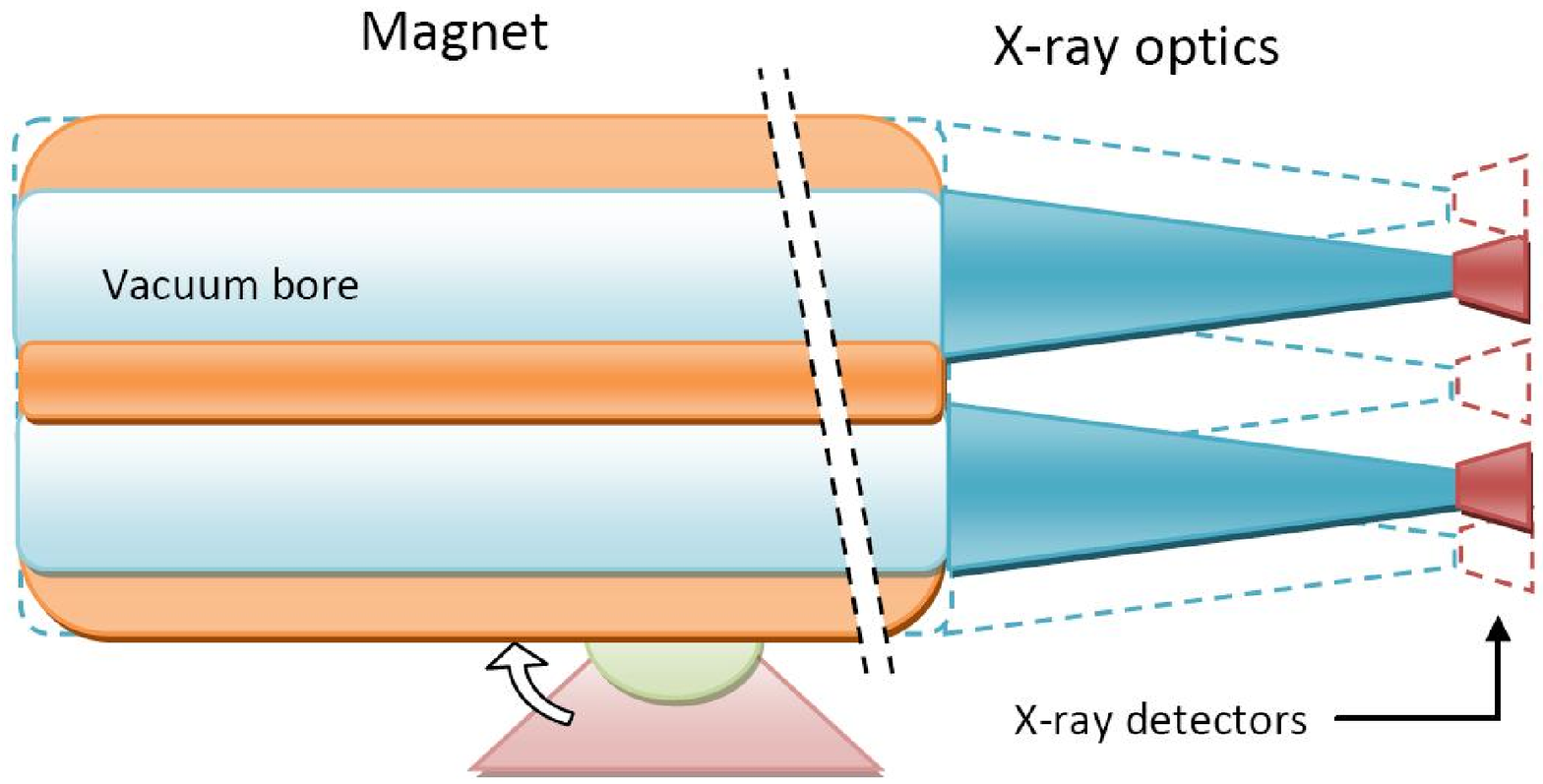}\hspace{2pc}%
\caption{\label{fig:sketch_NGAH} Possible conceptual arrangement
of the NGAH. On the left we show the cross section of the NGAH
toroidal magnet, in this example with six coils and bores. On the
right the longitudinal section with the magnet, the optics
attached to each magnet bore and the x-ray detectors.}
\end{figure}

In this subsection we work out the dependence of the sensitivity
to the axion couplings $\gagamma$ and $g_{ae}$ on each of the
experimental parameters of a NGAH in order to discuss the basis
for our proposed improvements. For this purpose, we define the
basic layout of an enhanced axion helioscope as one in which the
entire cross sectional area of the magnet is equipped with one or
more x-ray focussing optics and low background x-ray detectors.
This arrangement is schematically shown in figure
\ref{fig:sketch_NGAH}, in which we anticipate already a toroidal
design for the magnet as discussed later in section
\ref{sec:magnet}. The axion signal counts $N_\gamma$ and
background counts $N_b$ in such a layout can be written as:

\begin{equation}\label{ngamma}
N_\gamma \:\: \propto \:\: N^* \times g^4 \equiv
B^2 \: L^2 \: A \: \epsilon \: t \times g^4
\end{equation}

\begin{equation}
N_b \:\: = \:\: b \: a \: \epsilon_t \: t
\end{equation}

\noindent where $B$, $L$ and $A$ are the magnet field, length and
cross sectional area, respectively. The efficiency $\epsilon =
\epsilon_d \: \epsilon_o \: \epsilon_t$, being $\epsilon_d$ the
detectors' efficiency, $\epsilon_o$ the optics throughput or
focusing efficiency (it is assumed that the optics covers the
entire area $A$), and $\epsilon_t$ the data-taking efficiency,
i.~e.~the fraction of time the magnet tracks the Sun (a parameter
that depends on the extent of the platform movements). Finally,
$b$ is the normalized (in area and time) background of the
detector,  $a$ the total focusing spot area and $t$ the duration
of the data taking campaign. The relevant coupling constant $g$,
is $\gagamma$ for hadronic axions and $(\gagamma g_{ae})^{1/2}$
for non-hadronic axions.

Assuming that the measurement is dominated by backgrounds ($N_b>N_\gamma$)
but these can be estimated and subtracted through independent measurements (either by non Sun-tracking runs or by
using portion of detectors not exposed to the signal.\footnote{Assuming the detector is larger than the focus spot of the x-ray optic, the perimeter of the detector could be used for background determination.}), the discovery potential of the experiment depends upon
$N_\gamma/\sqrt{N_b}$
The sensitivity on the relevant coupling $g$ will be given by  $g \sim (N^*/\sqrt{N_b})^{-1/4}$.
It is useful to rewrite the previous expression in terms of
a figure of merit (FOM)
representing a growing relation with the ``merit'' of the experiment
\begin{equation}
f \equiv  \frac{N^*}{\sqrt{N_b}} = f_M \: f_{DO} \:f_T
\end{equation}
where we have factored the FOM to explicitly show the contributions from various experimental parameters: magnet, detectors and optics, and tracking (effective exposure time of
the experiment)
\begin{equation} \label{foms}
f_M  = B^2\:L^2\:A \;\;\;\;\;\; f_{DO}=\frac{\epsilon_d \:
\epsilon_o}{\sqrt{b\:a}} \:\:\:\:\:\: f_T=\sqrt{\epsilon_t\:t} \ .
\end{equation}

We stress that these expressions are obtained under two assumptions:
1) the axion-photon conversion is fully coherent, corresponding to the $L^2$
dependence shown in equation~\ref{foms}. 2) The exposure of the
experiment is such that we are in a gaussian regime, i.~e.~we have
at least $\agt 10$ background counts in the detectors.  For fewer
background events, background subtraction is not performed and the limit is
obtained in a different way than the one derived above.

As will be shown below, these FOMs clearly demonstrate the importance of the magnet parameters
when computing sensitivity of an axion helioscope. The CAST success has relied, to a large extent, on the
availability of the first class LHC test magnet which was recycled to become part of the CAST helioscope.
Going substantially beyond the CAST magnet's $B$
or $L$ is difficult, as 9 T is close to the maximum field one can
realistically get in current large-size magnets, while 10 m is a
considerable length for a structure that needs to be moved with
precision.
The improvement may come however in the cross section area, which
in the case of the CAST magnet is only $3\times10^{-3}$~m$^2$
Substantially larger cross sections can be achieved, although one
needs a different magnet configuration. It is an essential part of
our proposal that a new magnet must be designed and built
specifically for this application, if one aims at a substantial
step forward in sensitivity. We discuss in detail this issue in
section \ref{sec:magnet}, where we show that cross section areas
$A$ of up to few m$^2$ are feasible, while keeping the product of
$BL$ close to levels achieved for CAST.

Another area for improvement will be the x-ray optics. Although CAST has
proven the concept, only one of the four CAST magnet bores is
equipped with optics. The use of focusing power in the entire magnet
cross section $A$ is implicit in the FOM of equation~\ref{foms}, and
therefore the improvement obtained by enlarging $A$ comes in part
because a correspondingly large optic is coupled to the magnet.
Here the challenge is not so much achieving exquisite focusing or near-unity
reflectivity
(of course, the larger the throughput $\epsilon_o$ and the smaller the spot
area $a$, the better), but the availability of cost-effective x-ray
optics of the required size. This issue is discussed in detail in section \ref{optics}.

Finally, we need to discuss the x-ray detectors. CAST has enjoyed the sustained
development of its detectors towards lower backgrounds during its
lifetime. The latest generation of Micromegas detectors in CAST are
achieving backgrounds of $\sim$~$5\times10^{-6}$~\ckcs.
This value is already a factor 20 better than the backgrounds recorded during the
first data-taking periods of CAST. Prospects for reducing this
level to $10^{-7}$ \ckcs\ or even lower appear feasible and are discussed in section \ref{detectors}.

Although it has less impact on the sensitivity than the other factors, it is
also desirable to improve the tracking efficiency $\epsilon_t$.  The goal is to improve performance
 from the current value of 0.12 obtained with CAST to $\epsilon_t =$ 0.3--0.5.  This gain would help in gathering
exposure more quickly and shorten the time required for the experiment to move into the
non-zero background regime, where the above FOMs are applicable.  Higher efficiency is possible, provided that the design of the platform and magnet occurs in a coordinated fashion.

\begin{table}[t]
\begin{tabular}{ccccccc}
\hline  \textbf{Parameter} & \textbf{Units} & \textbf{CAST-I} & \textbf{NGAH 1} & \textbf{NGAH 2}& \textbf{NGAH 3}& \textbf{NGAH 4}\\
\hline \\
 $B$           & T         & 9               & 3       & 3     & 4     & 5     \\
 $L$           & m         & 9.26            & 12       & 15     & 15     & 20    \\
 $A$           & m$^2$    & 2 $\times$ 0.0015   & 1.7   & 2.6 & 2.6 & 4.0  \\
                                                                    \\
\hline
 $f_M^*$         &           & 1              & 100     & 260    & 450   & 1900  \\
                                                    \\

 $b$             & $\frac{10^{-5}\, \rm c}{\rm keV\, cm^2\, s}$ & $\sim 4$ & $3\times10^{-2}$ & $10^{-2}$ & $3\times10^{-3}$ & $10^{-3}$\\
 $\epsilon_d$  &           & 0.5 -- 0.9        & 0.7   & 0.7   & 0.7   & 0.7   \\
 $\epsilon_o$  &           & 0.3               & 0.3   & 0.3   & 0.6   & 0.6   \\
 $a$             &  cm$^{2}$ & 0.15               & 3   & 2   & 1   & 1   \\
 \hline
 $f_{DO}^*$      &           & 1               & 6   & 14   & 40   & 40   \\
                                                                        \\
 $\epsilon_t$  &           & 0.12              & 0.3   & 0.3   & 0.5   & 0.5   \\
 $t$             & year      & $\sim 1$          & 3     & 3     & 3     & 3     \\
 \hline
 $f_T^*$         &           & 1               & 2.7     & 2.7     & 3.5     & 3.5     \\                                                                        \\
 \hline
 $f^*$           &           & 1               & $1.6 \times 10^3$     & $9.8 \times 10^3$  & $6.3 \times 10^4$      & $2.7 \times 10^5$     \\
 \hline \hline
\end{tabular}
\caption{\label{scenarios} Values of the relevant experimental
parameters representative of CAST-I, as well as to the four
possible scenarios for a future NGAH referred in the text for
which the sensitivity is calculated. Numbers shown for the figures
of merit are relative to CAST-I, i.~e.~$f^* = f / f_{\rm CAST}$,
and are approximate.}
\end{table}

\begin{figure}[t] \centering
\includegraphics[height=7.5cm]{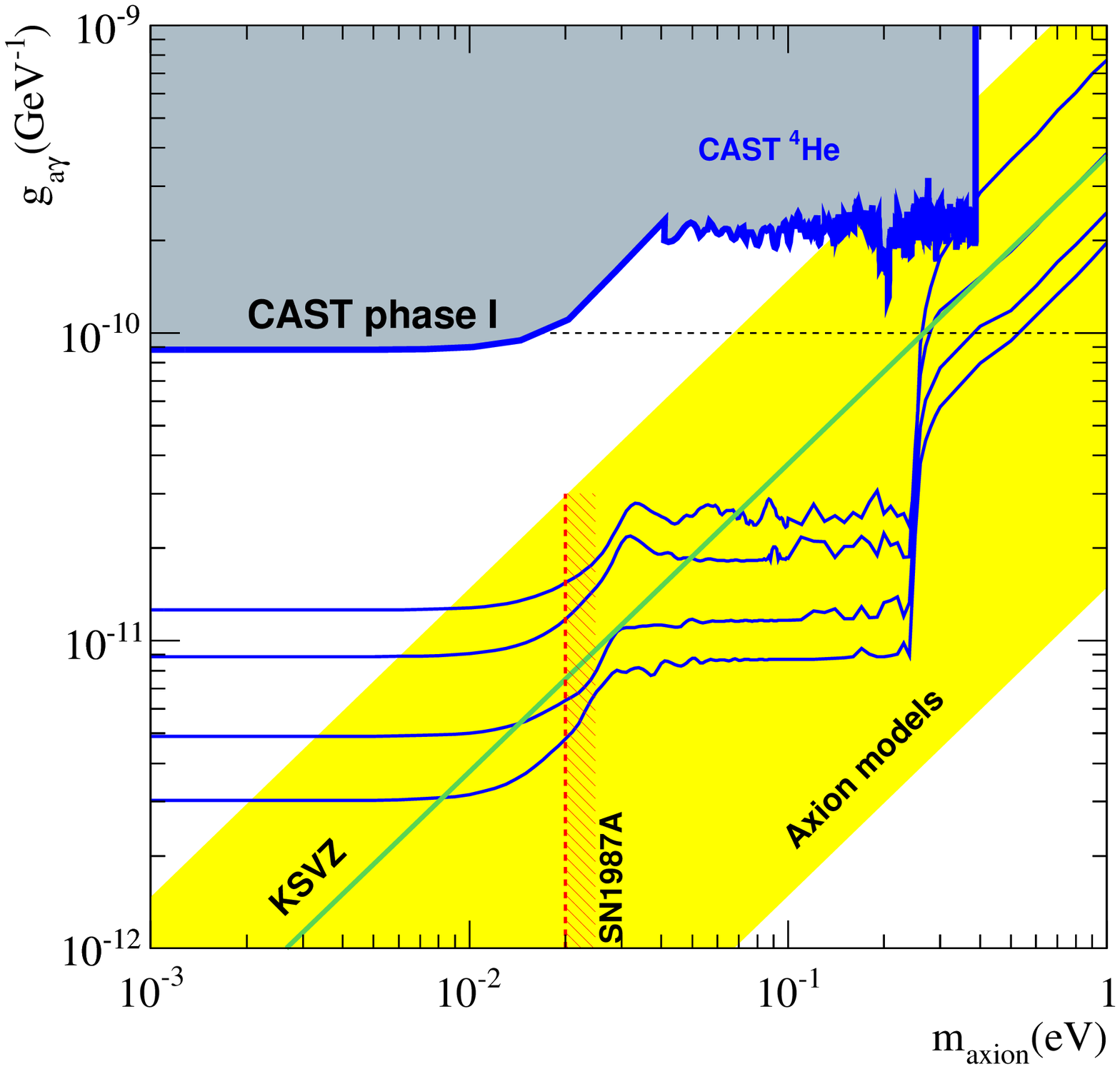}
\includegraphics[height=7.5cm]{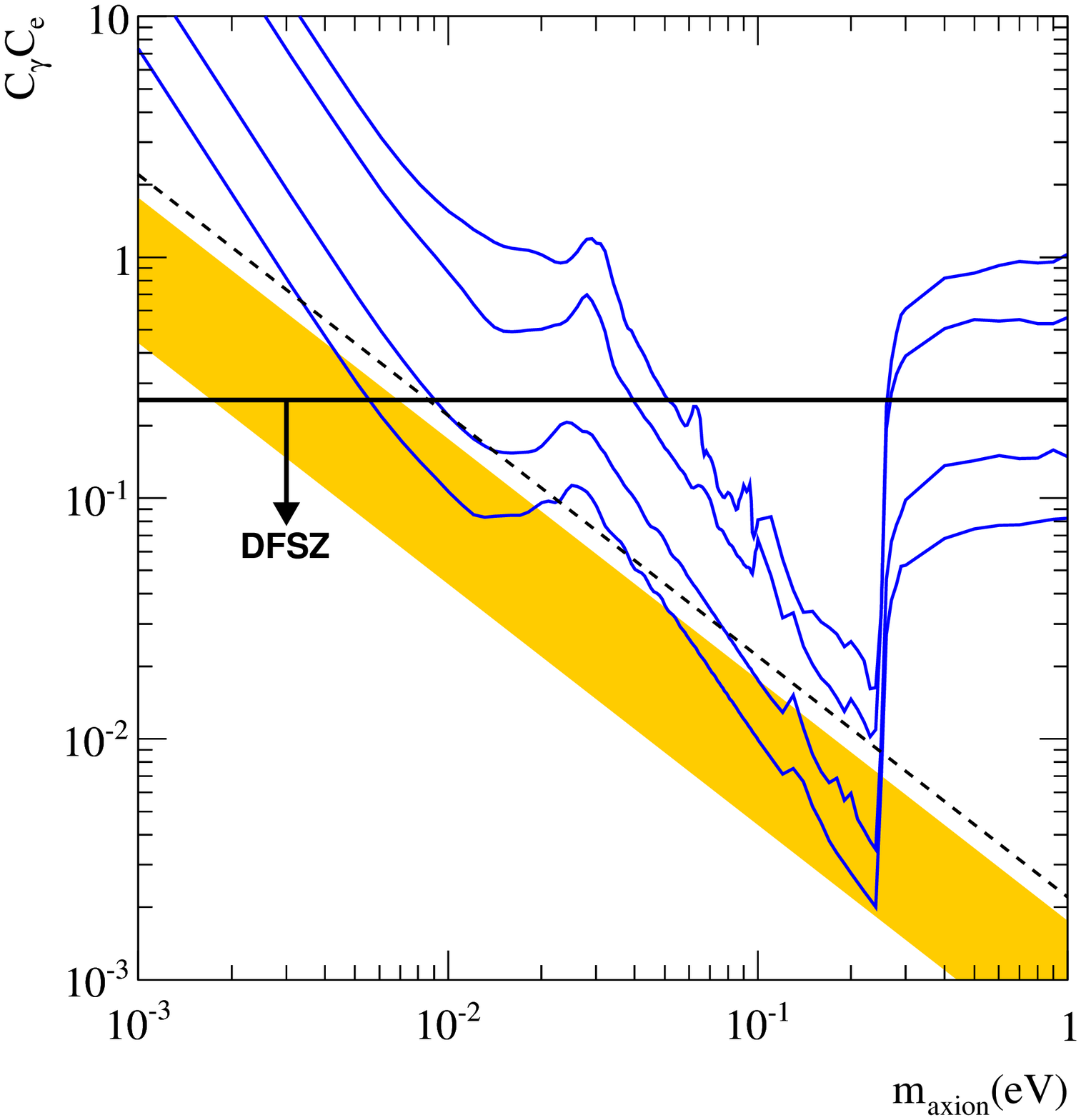}\hspace{2pc}%
\caption{\label{fig:scenarios} LEFT: The parameter space for hadronic axions and ALPs. The CAST limit, some other limits, and the range of PQ models (yellow band) are also shown. The blue lines indicate the sensitivity of the four scenarios discussed in the text and table 1.
RIGHT: The expected sensitivity regions of the four same scenarios in the parameter space of non-hadronic axions with both electron and photon coupling. In GUT
models $C_\gamma$ is fixed to $0.75$ and we show the bound on the
electron coupling ($C_e$) from red giants (dashed line along the diagonal) and the
region motivated by WD cooling (orange band).  DFSZ models lie
below the horizontal line $C_\gamma C_e< 0.25$.}
\end{figure}

The improvements suggested above could lead to sensitivities, in
terms of detectable signal counts, up to 10$^6$ better than CAST,
which corresponds to  $1.5$ orders of magnitude in $g$, as seen
from the FOMs of equation~\ref{foms}. In order to add fidelity to
our estimates, we have fully computed sensitivity plots for four
possible instantiations, and their associated experimental
parameters, of an enhanced axion helioscope. These four scenarios
are described in table \ref{scenarios} and include several
combinations of values ranging from less to more optimistic
assumptions and represent different degrees of success for the
improvements in the magnet, detectors and optics previously
mentioned.  Table \ref{scenarios} also contains data for CAST-I,
an experimental configuration that represents the current
state-of-the-art.

The computed sensitivities of each of the four NGAH scenarios are represented by the family of blue lines in figure~\ref{fig:scenarios}, both for hadronic axions (left) and non-hadronic ones (right). These calculations were performed by means of a Monte Carlo simulation of background counts,
computation of likelihood function and subsequent derivation of the 95\% upper limit assuming no detected signal. They include two data taking campaigns for each of the scenarios: one three years long performed without buffer gas (analogous to CAST I), and another three years long period with varying amounts of $^4$He gas inside the magnet bore (analogous to CAST II, although without the need to use $^3$He). This second phase is responsible for the step in the sensitivity line from mass of $\sim$0.05 eV up to 0.25 eV. This range is given by the gas density range chosen for this calculation of 0 to 1 bar of $^4$He at room temperature. Of course, a shorter density range could be chosen, thus allowing for a mass scan correspondingly shorter but more sensitive in $\gagamma$.
In general, the NGAH sensitivity lines go well beyond current CAST sensitivity for hadronic axions and progressively
penetrate into the decade $10^{-11}$--$10^{-12}$ GeV$^{-1}$, with
the best one approaching $10^{-12}$ GeV$^{-1}$. They are sensitive
to realistic QCD axion models at the 10 meV scale and exclude a good
fraction of them above this. For non-hadronic axions, the NGAH sensitivity lines penetrate in the DFSZ model region, approaching or even surpassing the red-giant constraints. Most relevantly, the NGAH 3 and NGAH 4 scenarios start probing the region of parameter space highlighted by the cooling of WDs.

\section{Magnet} \label{sec:magnet}

The previous analysis corroborates the importance of the magnet
for a competitive axion helioscope. As previously anticipated, in
order to achieve the stated step forward in sensitivity, the
design and construction of a new magnet is mandatory. Of course,
this must be done with the FOM for an NGAH in mind already at
design time. The latest magnet technology allows for the magnetic
strength and length to be improved with respect to CAST. However,
the needed margin for the required improvement in the magnet FOM
can still not be reached. Therefore, the magnet's aperture is the
only parameter left that can be significantly enhanced and thus we
shall base our possible magnet design by concentrating on it. We
must stress that thanks to the use of x-ray optics at the end of
the magnet bore, the enlargement of the magnet aperture does imply
an enhancement of the expected signal without necessarily implying
an increase of background. Indeed, as was shown before, the
overall FOM of the axion helioscope goes directly proportional to
the magnet bore area $f \propto A$ (which means that the
sensitivity to the coupling constant goes as $g_{\alpha\gamma}
\propto A^{-1/4}$). Needless to say, for this relation to hold,
one assumes the optics size is enlarged accordingly to couple the
magnet bore down to the stated focal spot size. It should also be
noted here that the magnetic field $B$ in an axion helioscope
magnet must be perpendicular to the longitudinal (axion incoming)
direction. More correctly, only the perpendicular components of
$B$, with respect to the axion beam momentum, will contribute to
the conversion probability.

%
Accelerator dipole magnets, like the one CAST is currently using, 
have additional design constraints that are not required by a NGAH
use, with the most important constraint being the extraordinary
quality of the magnetic field (i.~e.~the field is required to be
extremely uniform within the accelerator's aperture). Moreover,
accelerator type magnets cannot reach  apertures wide enough to
improve the magnet FOM significantly. For example, a CAST like
magnet with a 9 T magnetic field and 9.26 m length will need a 620
mm aperture, which is clearly not achievable in the near future,
in order to improve the relative FOM by a factor of just 100.
However, by considering different designs of detector
magnets,\footnote{following initial suggestions by L. Walckiers}
e.~g.~the ones of the ATLAS or the AMS experiments, which are
characterized by a very large volume and a lower field (compared
to accelerator magnets), it seems feasible to reach the required
FOM, in particular regarding apertures of up to several meters
with rather intense fields. A complete feasibility study is
currently in progress to define the simplest magnet design that
satisfies the requirements of a NGAH and optimizes the FOM within
the use of current magnet technologies at CERN.

\begin{figure}[hb!]
    \begin{center}
    \includegraphics[scale=0.6]{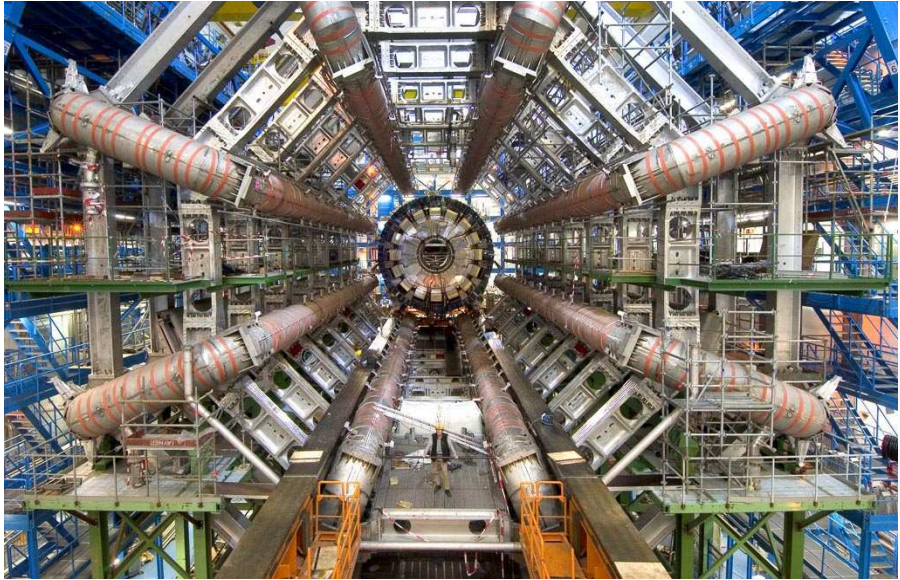}
    \end{center}
    \caption{The barrel toroid of the ATLAS experiment at CERN. The huge dimensions of the magnet can be appreciated by a comparison to the man standing at the bottom of the photo. The NGAH magnet's volume will be about 1--2 \% of this enormous magnet. Courtesy of the ATLAS experiment. } \label{fig:bar}
\end{figure}

\begin{figure}[t]
    \begin{center}
    \includegraphics[width=15cm]{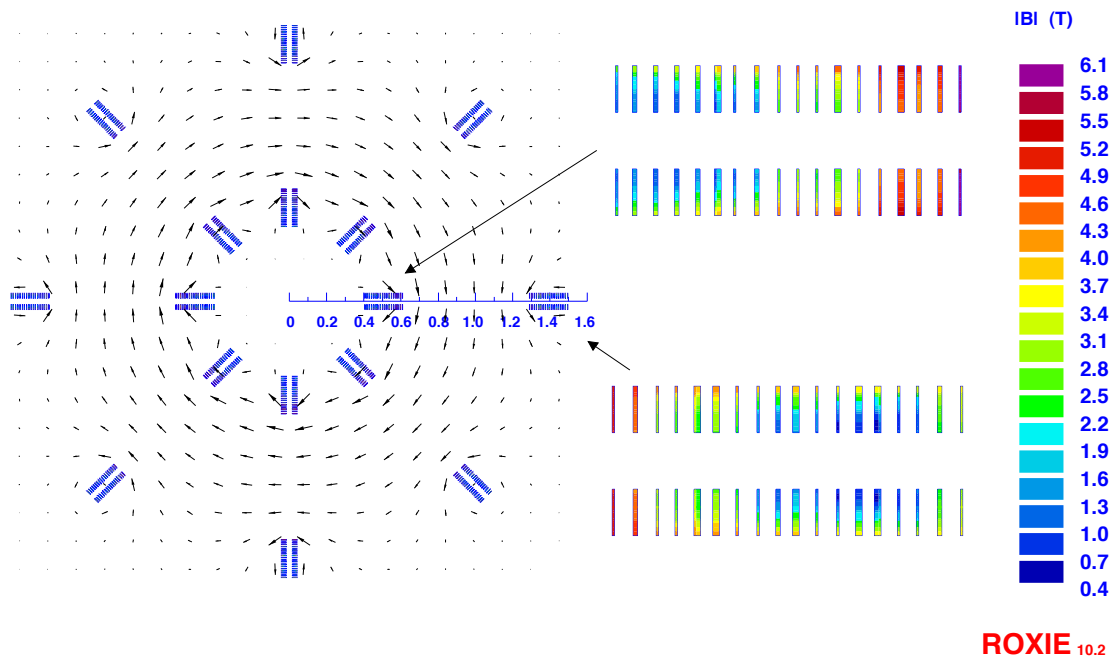}
    \end{center}
    \caption{An example for a possible toroidal NGAH magnet design. The cross-section of the toroidal magnet with 8 racetrack coils is shown on the left side of the figure and the modulus of the field inside the coils is represented on the right side, where a zoom of the inner (upper right hand side) and outer (lower right) coils is shown. In this possible design, the coils have a double layer geometry with 18 turns in each layer. The peak field is on the inner coil's internal side (with respect to the aperture) and is 6.1 T. The calculation was done with the CERN field computation program ROXIE 10.2.} \label{fig:tor}
\end{figure}

The ATLAS experiment at CERN is using an enormous central toroid
magnet \cite{ATLAS}, known as the barrel toroid, of 25.3 m in
length with 20.1 m and 9.4 in outer and inner diameters,
respectively (see figure~\ref{fig:bar}) . This toroid has a peak
field of 3.9 T at the coils which generate an average field of
about 0.8 T in the useful aperture (that is, the aperture that
would have been used for solar axions search) for a current of
20.5 kA. The NGAH can rely on these numbers and the barrel toroid
design in order to scale it down and optimize it for axions
search. Since the useful diameter for the optics detector is not
more than 1 m, the NGAH has the advantage of having a smaller
width and hence maintaining a higher useful field in the aperture.
First considerations seem to favor a configuration in which 8
vacuum bores, of relatively large size (0.5--1 m diameter), are
placed between the coils and are available to couple optics and
detectors. This configuration is demonstrated in
figure~\ref{fig:tor}, where a cross-section of the geometry is
shown. The magnetic field in the bores, although not homogeneous,
is largely perpendicular to the axion directions. With this type
of configuration, it appears possible to reach a magnet length
similar or somewhat longer than CAST (15--20 m), with B peak
fields not much less (about 6 T) and a total cross sectional
magnetic area of around 1--3 m$^2$. The average field in the
aperture (i.~e.~the vacuum bores) is about 2.5--3 T, thus
providing a FOM in the range of 100--350 relative to CAST,
already reaching or surpassing the values stated in scenarios NGAH-1 and NGAH-2 of table \ref{scenarios}.

\begin{figure}[ht!]
    \begin{center}
    \includegraphics[width=10cm]{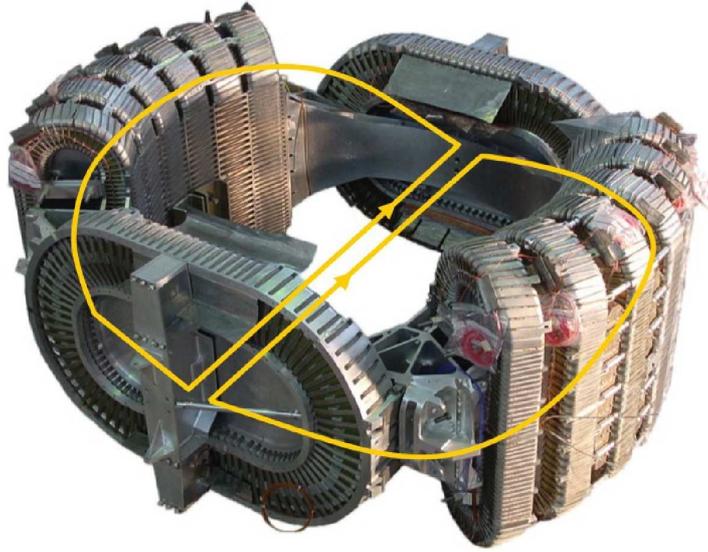}
    \end{center}
    \caption{The AMS superconducting magnet. The two largest coils generate the dipolar field while the 2 $\times$ 6 shielding coils close the magnetic flux and reduce considerably the fringe fields. Source: http://www.ams02.org/what-is-ams/tecnology/magnet/scmagnet/} \label{fig:ams}
\end{figure}

As mentioned above, also in this class of magnets one uses the
super-conducting shielded dipole magnet design, which was designed
by the AMS mission \cite{ams}. This NGAH design will have a dipole
field in its center, where the dipole is surrounded by an 8 coils,
semi-toroidal, geometry (see figure~\ref{fig:ams}). The dipole
bore can contain 6--8  apertures with an average field of about
1.5 T, while the peripheral shielding coils give additional 2
apertures with an average field of 2.5 - 3 T and 4 apertures (when
using 4 shielding coils and not 6 as in the original AMS design)
with an average field of 2.5 T. Overall, this geometry will not
yield a higher FOM than the one that can be gained with the
toroidal design and will also have the disadvantage of using more
cable, which increases the overall costs of the NGAH. Moreover,
another disadvantage of the AMS geometry is that it sustains
higher stress than the ATLAS geometry since the toroidal geometry
is self supported thanks to its symmetry. The bigger stress serves
as an additional limitation on the maximal current and hence on
the magnetic field.

Another option for the NGAH will be to consider a solenoidal
magnetic field. These kind of magnets have the advantage of being
the easiest to design and manufacture. However, since in a CAST
like experiment the axion beam has to be perpendicular to the
magnetic field, the solenoid must be transparent to x-ray photons,
which limits the magnetic field strength and the radius of the
solenoid and makes achieving the FOM goal very difficult. On top
of that, a solenoid magnet with the parameters needed for the NGAH
(i.~e.~large diameter and very high field) will suffer from very
large fringe fields which will restrict the possibility for easy
approach and access to detectors, optics and cryogenics.

The use of a new and more advanced superconductor
(SC) such as Nb$_3$Sn has also been considered. Nb$_3$Sn may increase the
magnetic peak field up to 15--16 T in an accelerator type magnet
(for the same amount of SC). However, such an increase will double
the stress applied to the coils, which is already close to the
limit at the 9 T dipole. In addition, this material is about 5
times more expensive than NbTi and, moreover, these magnets are
still in a R\&D stage. The use of Nb$_3$Sn has also a limitation
since it is strain sensitive and very brittle. It practically
ceases operating when the stress is above 150 MPa.
Provided these issues are solved in the future, this material would however represent a large improvement beyond our most optimistic assumptions.

The major efforts when coming to engineer the NGAH magnet, will
focus on the mechanical structure, cryogenics and (quench)
protection of such a system. Since the required increase of the
present FOM is of a large factor, which will be challenging to
achieve, the new design will have to stretch the limits of the
design factors (such as operating current, operational margin,
cable design and inductance). Nonetheless, it will be more
efficient to follow known designs and by that reduce the need for
building and designing new tooling and assembly machines.

\begin{figure}[ht!]
\begin{center}
\includegraphics[width=7.5cm]{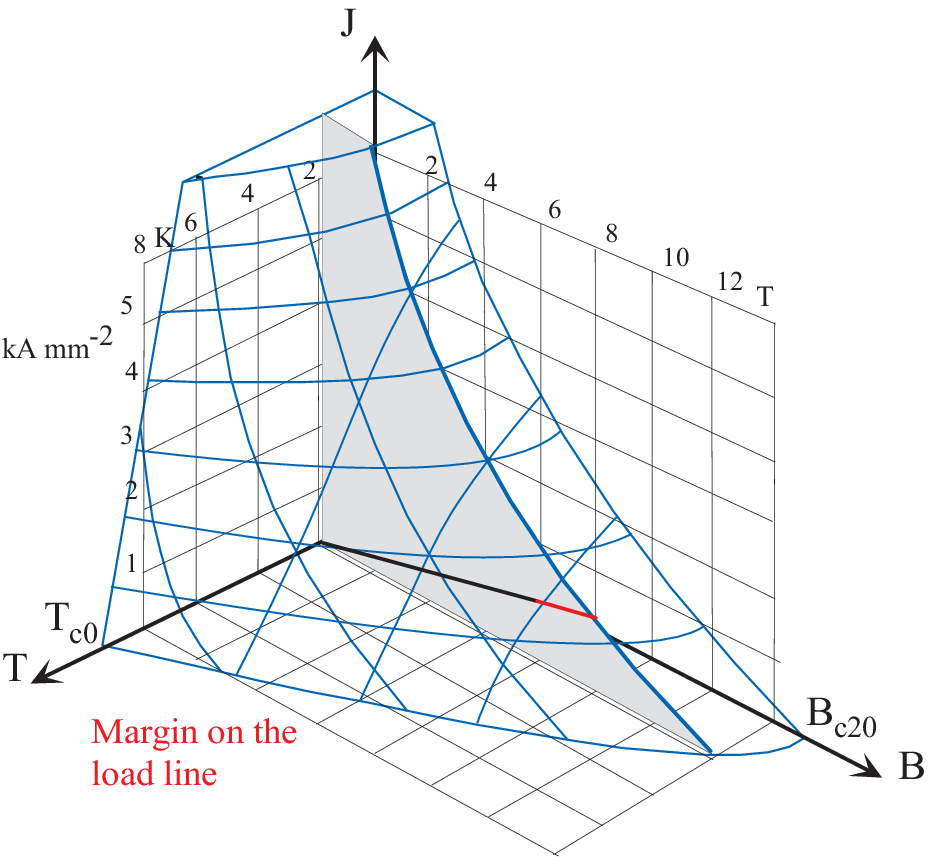}
\includegraphics[width=7.5cm]{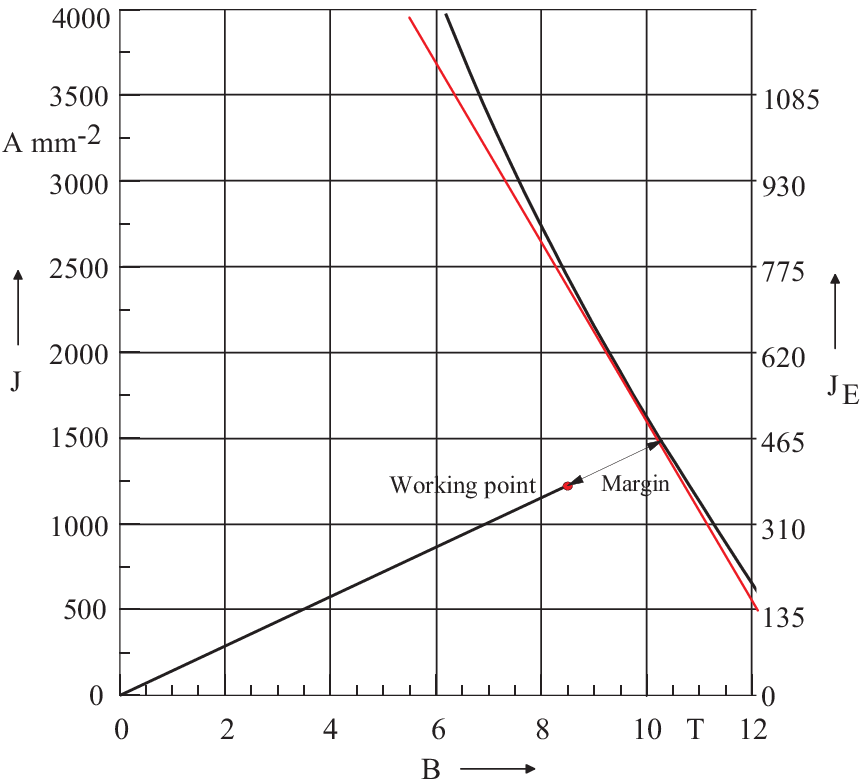}
\caption{\label{fig:cs} Left: Critical surface of NbTi
superconductor. Also shown are the load line (continuous straight
line, divided to two parts). The operational margin at a constant
temperature is the red portion of the load line, while the working
point is at the end of the black portion of the load line. Right:
Critical current density of NbTi at 1.9 K (black line), together
with the linear approximation for the critical current density
(red line), the load line and the working point. These images
represent data taken from the LHC main dipoles.}
\end{center}
\end{figure}

In this context, it is important to emphasize that the NGAH magnet
requires a very large aperture while still maintaining the highest
possible magnetic field. To understand the difficulties in
achieving this, two definitions are required: The so-called
operational margin of the magnet and the magnet's load line  (see
figure~\ref{fig:cs}).

For a superconducting magnet, the magnetic field is limited by the
critical surface, which is determined by the properties of the
superconducting material. This means that for a given temperature
and current density there is a critical magnetic field limiting
the superconducting performance. Hence, for the sake of proper
magnet operation, the magnetic field should be low enough to avoid
frequent quenching but, at the same time, for the efficiency and
the purpose for which the system is being built in the first
place, the magnetic field should be as high as possible. This
choice of the operating envelope of the magnet determines the
operational margin of the magnet.

The operational margin is defined by means of the load line: For a
given configuration of the magnet (at a constant temperature), the
Biot-Savart law gives the (linear) relation between the current
density $J$ and the magnetic field strength $B$. This relation
yields a straight line in the $(B, J)$ phase space. The portion of
this line, which extends from the origin of the $(B, J)$ space to
the critical surface is called the ``load line'' of the magnet. In
the magnet designers' jargon, it is common to refer to the
operational margin by the so called percentage on the load line.
For example, the operational margin of the LHC main dipoles is set
at 20$\%$ on the load line \cite{Bruning:2004ej}. This expression
means that the magnet's operating values, namely the current
density and the magnetic field, are those given by the point
(called the magnet's working point) in the $(B, J)$ phase space
which will mark 0.8 of the magnet's load line length, for a given
temperature (1.9 K for the LHC). The smaller the operational
margin is, the closer the magnet is to its quench point.

Most detector type magnets usually work at lower fields and hence
have a relatively large operational margin. However, as mentioned,
the NGAH magnet will have to sustain the highest possible fields.
Thus, the operational margin will inevitably be reduced and the
NGAH will have to combine the protection techniques commonly used
for both detector and accelerator magnets. Consequently, the
protection of the NGAH magnet may be the most challenging part of
the design.

An efficient magnet design will be the one which yields a load
line with a slope as small as possible since the smaller the load
line's slope is, the lower the current density needed to generate
a specific magnetic field. A possible way to increase the
operational margin, for a given temperature, is by adding more SC
to the magnet while at the same time reducing the current density.
Thus, by increasing the number of current sources, the same field
can be maintained for a lower current density. This can be done in
two ways: the straight forward way will be to increase the number
of turns in the coil, or more simply, using more cable. However,
additional turns make the magnet's coil, and hence the cold mass
(i.~e.~the coils and their supporting structure) as well, bigger.
Therefore, the resulting magnet will have an aperture bigger than
the area that can be covered by the optics and this might result
in a reduced efficiency (depending on the gain in the magnetic
field). Another way to increase the SC amount is to use larger
strands in the cable and by that not affecting considerably the
geometry of the magnet.

It is important to notice, however, that even increasing the
amount of SC in the coils has a limited influence on the magnet's
performance, since, at some point, adding more SC to the coil will
increase the cost without adding significantly to its
capabilities. Moreover, the magnetic field is always constrained
by the critical field at zero current density, which is an
internal property of the superconducting material.

For example, for a dipole magnet as the one used by CAST, the
relation between the modulus of the magnetic field $B$ to the
critical current density $J_c$ of the SC is given by
\cite{stephan}

\begin{equation}
\label{btoj}
    B = \frac{\mu_0}{2} \lambda_{\rm tot}J_c W ~,
\end{equation}

\noindent where $\mu_0$ is the vacuum permeability, $\lambda_{\rm
tot}$ is the total superconducting filling factor (the ratio
between the engineering current density to the critical current
density) and $W$ is the width of the coil. Using the linear
approximation of the critical current density $J_c$ for NbTi

\begin{equation}
J_c = d(\tilde{B}_{c2} - B)~,
\end{equation}

\noindent valid for the high field region (i.~e.~for magnetic
fields larger than 5 T at 1.9 K and 2 T at 4.2 K) and for
$\tilde{B}_{c2}>B$, where $d = -\frac{dJ_c}{dB}|_{\tilde{B}_{c2}}$
is the negative slope in the high field region of the critical
surface at constant temperature and $\tilde{B}_{c2}$ is the
critical field at zero current density according to the fit
\cite{PhysRevSTAB.10.112401}, we get the relation between the
modulus of the magnetic field $B$ and the critical field
$\tilde{B}_{c2}$

\begin{equation}
\label{btobc}
    B = \frac{\mu_0}{2} \lambda_{\rm tot} (\tilde{B}_{c2} - B)W d ~.
\end{equation}

By obtaining an expression for the width of the coil from the
latter

\begin{equation}
    W = \frac{2B}{\mu_0 \lambda_{tot}d(\tilde{B}_{c2}-B)}~,
\end{equation}

\noindent one immediately notices that $W \rightarrow \infty$ as
$B \rightarrow \tilde{B}_{c2}$. Therefore, when increasing the
coil's width $W$ the magnetic field $B$ will rise and at the same
time the critical current density ($J_c = (\tilde{B}_{c2} - B)d$)
will have to be decreased (in order to stay below the critical
surface). By linearity of the Bio-Savart law, this implies that
the slope of the load line will be smaller.

From Eq. (\ref{btobc}) we can also derive an explicit expression
for the magnetic field

\begin{equation}
\label{ }
    B = \frac{\tfrac{\mu_0}{2} \lambda_{tot} W d}{1 + \tfrac{\mu_0}{2}\lambda_{tot}W d}\tilde{B}_{c2} ~,
\end{equation}

\noindent from which it is seen that the relations $\tilde{B}_{c2}
> B$ and $\partial B/\partial W > 0$ hold for any $W > 0$. Hence,
adding current sources to the coil will loose its efficiency at
some point since the $B(W)$ curve is asymptotically approaching
the limit value $\tilde{B}_{c2}$. Similarly, the same conclusions
of the last paragraph hold for detector type magnets as well.

Work is still ongoing to further define the geometry, dimensions
and final magnetic field strength, as well as the technical issues
and cost. However, the preliminary considerations exposed above, still not exhaustive, indicate that the toroidal configuration, inspired by ALTAS, in a favorable choice to at least achieve the magnet parameters listed in scenarios NGAH-1 or NGAH-2 of table~\ref{scenarios}. These scenarios represent conservative assumptions for our sensitivity prospects. It is not unrealistic to assume that a more detailed optimization study could yield an improved set of parameters as the one represented by scenario NGAH-3 or, more optimistically, NGAH-4. Provided the issues regarding the use of new SC materials like Nb$_3$Sn are solved in the future, these optimistic scenarios or even better ones could certainly be achieved .

The toroid design seems also to be the simplest and
cheapest way to achieve those FOMs. Also, we can base most
of the design on the existing and proven technology and the R\&D
that was carried out in order to be used in the ATLAS magnet.
Moreover, there are certain points of this design option that
represent important qualitative advantages with respect to the
current CAST experience:

\begin{itemize}
  \item The CAST magnet needs a heavy iron yoke around the bore and coils, in which to let the field lines to close, and thus preventing the field to leak out of the magnet (fringe field). The coil arrangement in a toroidal geometry is such that they lead the field on a close compact path and there is no need for iron yoke. Hence, almost all the Ómagnetic volumeÓ produced by the magnet can effectively be used for axion conversion. This is not the case in the CAST magnet, in which part of the magnetic flux is lost for axion-detection purposes inside the iron yoke. This leads to a more efficient use of the magnet strength.

  \item For the same reason (no need for an iron yoke), the weight of the magnet compared to its volume is much lower than in the current CAST magnet. For example, a toroid magnet will weigh about 10 times more than the CAST magnet, but will have an effective useful volume (i.~e.~volume used for data taking) of 700--1100 times (depending on the length) more than the twin dipoles of CAST.

    \item The cryogenics to cool down the superconducting coils are confined around the coils themselves, independent of the vacuum pipes (i.~e.~the magnet's apertures) which lie in between the coils, thus leaving them at room temperature. This arrangement, unlike the CAST one in which the magnet bore was cooled down to cryogenic temperatures together with the coils, results in a more practical operation in several aspects: no big cryostat enclosing all the magnet, easier access to the magnet bores (pumps, sensors, etc.), no cryogenic pumping effects in the vacuum system, no need to use $^3$He in a possible second phase with buffer gas (because $^4$He at room temperature can go to the required pressures, while in CAST $^3$He is needed as $^4$He would condense at 1.8~K).

\end{itemize}

\section{X-ray optics} \label{optics}

\subsection{Existing technologies}

X-ray optical designs are plentiful and rely on several different
phenomena including total external reflection (e.~g.~Kirkpatrick
Baez optics \cite{kb48}), refraction (e.~g.~aluminum-based
compound lenses \cite{sksl96}) and diffraction (e.~g.~Fresnel zone
plates \cite{drg+99}). For an axion helioscope, there are three
primary drivers for selecting the appropriate type of optic:
high-efficiency in the 1--10 keV energy range (0.3--5 keV for the
non-hadronic axions), a pupil entrance whose area is well-matched
to the area of the magnet bore and a solid-angle acceptance
greater than the $\sim$3 arcmin (0.87 mrad) extent of the solar
core where axions are produced. After considering these top-level
requirements, it is clear that reflective x-ray optics are the
obvious choice for the NGAH.

For more than forty years, reflective x-ray optics have been
continually refined and employed for either high-energy
astrophysics, a field where satellite-based telescopes are built
with multiple ``nested mirrors'' (a few to hundreds of mirror
layers) to achieve large geometric area (hundreds to thousands of
cm$^{2}$), or for x-ray light sources (e.~g.~synchrotrons or
free-electron lasers), where high-spatial resolution or minimal
wavefront-distorting optics are built with one or two reflective
elements with small apertures (typically a few mm$^{2}$ or
smaller).  At these photon energies, the total external reflection
of light occurs at very shallow incident angles ($<1 \deg $) and
so the terms glancing- and grazing-incidence are synonymous with
reflective x-ray optics that operate above 1~keV.

For light-source applications, the current state of the art in
reflective x-ray optics involves trying to achieve extremely high
spatial resolution ($\sim$ 10~nm) with focusing systems
\cite{mhk+10} or extremely smooth surfaces (figure errors less
than 1~$\mu$m) for relay systems \cite{pbm+07}.  The x-ray beams
produced by light sources are intrinsically small (hundreds of
$\mu$m$^{2}$ in cross-sectional area) and have low divergence (a
few to several $\mu$rad), so a single mirror element (for relays)
or two mirrors (for focusing systems) are often sufficient.  These
optics can often have additional requirements, like being able to
withstand heat loads of up to thousands of watts or being
compatible with ultra-high vacuum conditions ($<10^{-9}$~Torr), so
the final optical system can include integrated and complex
engineering features, which can significantly increase costs.
Thus, as impressive as these technologies are, they are not
well-matched for building large-area optics capable of accepting a
(relatively) large divergence source.

Instead, the appropriate choice is the nested designs
utilized by the astronomy community, specifically  those based on the ideas
of Wolter \cite{wolter52}.  The Wolter I is the most commonly used system, and
consists of a surface of revolution generated from a parabola for the initial reflection
and a surface of revolution generated from a hyperbola for the secondary reflection.
(Two reflections allows the Abbe sine rule to be nearly satisfied and allows off-axis
imaging with acceptable levels of aberration.)  The Wolter I prescription has the distinct
advantage that successively smaller radii shells can be placed inside one another or
``nested'', much
the way wooden Russian dolls fit inside each other.  The paraboloid and hyperboloid
shapes can be approximated by truncated cones \cite{ps85}.  Although on-axis resolution is
sacrificed, these so-called conical approximation or Wolter-I-like designs have good
off-axis performance and can be considerably less complex and less expensive to fabricate.

For astrophysics, the state-of-the-art x-ray optics (or
telescopes) are flying on NASA's {\it Chandra X-ray Observatory}
\cite{wtvo00} and ESA's {\it XMM-Newton} \cite{jla+01}.  {\it
Chandra's} single telescope, consisting of four nested layers
ground from monolithic Zerodur blanks, has exquisite spatial
resolution (0.5 arcsec half-power diameter) and modest effective
area (800 cm$^2$ at 1 keV), while {\it Newton's} three telescopes,
each consisting of 56 nested shells produced via replication, have
modest spatial resolution (15 arcsec half-power diameter) and
large effective area (a combined 4500 cm$^2$ at 1 keV). The
impressive telescope performance of these major observatories came
at high prices (700M USD for {\it Chandra}) and (100M EUR for {\it
Newton}), and today the astronomy community is trying to develop
lower-cost alternatives for the substrates.

\subsection{Technologies under development for astrophysics}
In this section, we discuss different approaches for
fabricating telescope substrates.  For each technology, we give
a brief description and cite examples of telescopes
that rely on it.  Broadly speaking, telescopes can be classed
into two groups that depend on how they are assembled.
Segmented optics rely on several individual
pieces of substrates to complete a single layer.  (The appropriate
analogy is the way a barrel is assembled from many individual staves.)
Integral-shell optics are just that:  the hyperbolic or parabolic
shell is a single monolithic piece.

\subsubsection{Segmented optics:  rolled aluminum substrates.}
Telescopes formed from segmented aluminum substrates were first
utilized for the broad band x-ray telescope (BBXRT) that flew on
the Space Shuttle in 1990 \cite{ps85}.   Later missions that used
the same approach included {\it ASCA}~\cite{sjs+95}, launched in
1993, SODART~\cite{cmh+97}, completed in 1995 but never launched,
the hard x-ray, balloon-borne InFoc$\mu$s~\cite{oto+02}, flown in
2004, Astro-E~\cite{hie+01}, destroyed on launch in 2000, and {\it
Suzaku}~\cite{ssc+07}, launched in 2005. Aluminum substrates will
also be used for the soft and hard x-ray telescopes
\cite{ssoh10,kaf+10} on the upcoming JAXA Astro-H mission,
scheduled for launch in 2014.

\subsubsection{Segmented optics:  glass substrates}
Although using glass substrates for an x-ray telescope was
explored as far as back as the 1980s \cite{labov88}, it was not
fully realized until the construction of HEFT \cite{hbb+00} in the
mid 2000s.  HEFT, flown from a balloon in 2005,  had three, hard
x-ray telescopes, each consisting of as many as 72 layers.  HEFT
is the pathfinder for the {\it NuSTAR}, a mission scheduled for
launch in 2012 \cite{hab+10}.  Each of {\it NuSTAR's} two
telescope consists of 130 layers, comprised of more than 2300
multilayer-coated pieces of glass \cite{hab+10b}.    Finally,
slumped glass is a candidate technology being developed by several
groups \cite{zab+10,wvf10,gbb+10} for the International X-ray
Observatory (IXO)~\cite{bookbinder10}, a mission being developed
in cooperation between NASA, ESA and JAXA.

\subsubsection{Segmented optics:  silicon substrates}
Another technology being pursued for IXO are pore optics, which
consists of silicon wafers that have a reflective coating on one
side and etched support structures on the other \cite{cga+10}.
Individual segments are stacked on top of each other to build
nested layers.  Prototype optics have been built and tested, but
there are no operational x-ray telescopes yet to use this method.

\subsection{Integral shell optics:  replication}
Replicated optics are created by forming the mirror,  usually a
nickel-based alloy, on top of a precisely figured and polished
mandrel or master.  The completely formed shell is separated from
the mandrel.  A mandrel is required for each unique layer. The
first mission that used replicated mirrors was EXOSAT
(\cite{dgc+81}, launched in 1983) and was followed by {\it
Beppo-SAX} (\cite{ccm+85}, launched in 1996), {\it ABRIXAS}
(\cite{eak+98}, launched in 1999), {\it XMM} and the balloon-borne
HERO experiment (\cite{res+99}, first flight in 2002). More
recently, replicated telescopes are being constructed for
eROSITA~\cite{pab+10}, an x-ray instrument on the
Spectrum-Roentgen-Gamma (SRG) satellite scheduled for launch in
2012 and FOXSI \cite{kcg+09}, a rocket-based solar instrument
scheduled for its initial flight in 2011.

\subsection{Integral shell optics:  monolithic glass}
For completeness, we mention telescopes formed from
monolithic pieces of glass.  {\it Einstein} (\cite{gbb+79}, launched in 1978), {\it RoSAT} (\cite{trumper83}, launched in 1980)
and {\it Chandra} are the three major missions that had these type
of telescopes.  Because of the cost and weight of the mirrors, no future
mission is expected to use this fabrication method. Table~\ref{tab:optics} summarizes the design parameters of many of the telescopes discussed in the previous text.

\begin{table}
\
\begin{tabular}{l l l c c c c c}
\hline \textbf{Mission} & \textbf{Design} & \textbf{Fabrication} & \textbf{F [m]} & {\boldmath $\rho$} \textbf{[mm]} & {\boldmath $\alpha$} \textbf{[deg]}   & \textbf{Layers} & \textbf{Ref}\\
\hline \hline
ABRIXAS        & Wolter I  & replication    & 1.60     & ~38$-$82~  & 0.33$-$0.72  & ~27  & \cite{eak+98}\\ 
Astro-H (hard) & cone      & seg. alum      & 12.0    & ~60$-$225  & 0.07$-$0.27  & 213 & \cite{kaf+10}\\ 
Astro-H (soft) & cone      & seg. alum      & 5.60     & ~60$-$225  & 0.14$-$0.54  & 203 & \cite{ssoh10,okm+08}\\ 
BeppoSAX       & cone      & replication    & 1.85    &  ~33$-$81~  & 0.26$-$0.62  & ~30 & \cite{ccm+85} \\ 
Chandra        & Wolter I  & monolithic     & 10.0    & 320$-$600  & 0.45$-$0.85  & ~~4  & \cite{woev95}  \\
eRosita        & Wolter I  & replication    & 1.60     & ~38$-$180  & 0.33$-$1.60  & ~54 & \cite{pab+10,eak+98} \\
HEFT           & cone      & seg. glass     & 6.00     & ~40$-$120  & 0.09$-$0.29 & ~72  & \cite{hbb+00}\\
NuSTAR         & cone      & seg. glass     & 10.2   & ~54$-$191  & 0.08$-$0.27  & 130 & \cite{hab+10}\\ 
SODART         & cone      & seg. alum      & 8.00     & ~80$-$300  & 0.15$-$0.54  & 143 & \cite{cmh+97}\\ 
XMM-Newton     & Wolter I  & replication    & 7.50     & 153$-$350  & 0.29$-$0.67  & ~58 & \cite{jla+01} \\ 

\hline
\end{tabular}
\caption{\label{tab:optics} Properties of x-ray telescopes made
for different observatories. $F$ is the focal length, $\rho$ is
the range of shell radii and $\alpha$ is the range of graze
angles. References for telescope parameters are given in the last
column.}
\end{table}

\subsection{Considerations for a NGAH}

When designing x-ray optics for a NGAH, several interrelated
factors must be considered. Of paramount importance is how optical
properties like efficiency and spot-size, key parameters for
computing the FOM, $f$, directly impact experimental sensitivity.
These performance characteristics will depend on manufacturing
technique and optical prescription. Both of these choices will
drive cost.  Additionally, the physical size of the optic will
influence the overall design of the infrastructure required for
the NGAH (e.~g.~tracking platform and the structure in which the
experiment will be housed) and thus will also influence costs.

It is beyond the scope of this paper to present the
results of a full design of an x-ray optic intended
for the NGAH.  In fact, the optimization of the optics
will be intimately linked to the magnet design and that of
the entire facility and, thus, cannot be completed until the
overall scope of NGAH is better defined.

Instead, we show it is feasible to obtain values of $\epsilon_{o}$
and $a$ consistent with those presented in table~\ref{scenarios}.
We start with the idea that there will be one telescope for each
of the magnet bores, as shown in figure~\ref{fig:sketch_NGAH}. For a total area
of 3.0~m$^{2}$, each telescope must have an entrance pupil of 399~mm.

The basic equations that govern the design of a Wolter x-ray telescope is
the relationship between focal length $F$, the radius of the shell $\rho$, and
the graze angle, $\alpha$:
\begin{equation}
F = \frac{\rho}{\tan 4 \alpha}
\label{eq:optic1}
\end{equation}
and the radius $r$ of the projected solar core, of
angular width $\omega \approx 3~{\rm arcmin}$, at the detector plane\begin{equation}
r = F \times \tan(\omega/2) \approx \frac{1}{2}F  \omega.
\label{eq:optic2}
\end{equation}
For a cone-approximation to a Wolter telescope, there is a small
modification to equation~\ref{eq:optic1} which we ignore below.

To compute $a$, we first increase the radius of the focused spot by 25\%
to account for imperfections in the mirror (e.~g.~figure errors)
that broaden the point spread function:
\begin{equation}
a = \pi [r(1+0.25)]^{2} = \pi \frac{25}{16} r^{2} \approx
9.3\times10^{-3} F_{\rm m}^{2} {\,\rm cm}^{2}, \label{eq:optic3}
\end{equation}
where $F_{\rm m}$ is the focal length in meters.

X-ray reflectivity of mirror coatings depends
on several material factors, including density,
optical constants and surface roughness.  Working at
graze angles at or below 1.0$^{\circ}$ ensures high
reflectivity for common coating materials like nickel
or gold.  Based on this knowledge, we consider two
telescope designs, one with a maximum graze angle
of 1.0$^{\circ}$ and another with a maximum graze
angle of 0.75$^{\circ}$.  According to equation~\ref{eq:optic1},
this translates to a focal length of 5.7~m and 7.6~m;
and according to equation~\ref{eq:optic3},
a focus area, per telescope, of 0.30~cm$^{2}$ and 0.53~cm$^{2}$.
Fixing the length of each mirror section at 300~mm and starting
with a maximum radius of 399~mm, we then generated
complete prescriptions for cone-approximation Wolter I telescopes.
We adopted design principles (i.~e.~spacing between layers, support
structures and gaps between individual segments) developed for segmented
glass substrate telescopes like HEFT and NuSTAR \cite{kcc+04}.
Table~\ref{tab:optics2} details the properties of the two designs.

\begin{table}
\centering
\begin{tabular}{l c c}
\hline \textbf{Parameter} & \textbf{Design 1} & \textbf{Design 2} \\ \hline \hline
F & 5.67 m & 7.58 m \\
$\rho$ & 100$-$399 & 100$-$399 \\
$\alpha$ & 0.25$-$1.00$^{\circ}$  & 0.19$-$0.75$^{\circ}$ \\
Layers & 95 & 123 \\
Substrate thickness & 0.3 m & 0.3 m \\
Geometric area [m$^{2}$] & 0.40 & 0.39  \\
Mirror coating & Ni & Ni \\
$r$~[mm] & 3.1 & 4.1 \\
$a$~[cm$^{2}$] & 0.30 & 0.53 \\
$\epsilon_{o}$ & 0.37$-$0.49 & 0.28$-$0.37 \\

\hline
\end{tabular}
\caption{\label{tab:optics2} Properties and parameters for two
NGAH telescope designs. The geometric area refers to the total
projected on-axis area of the telescope and includes losses due to
support structures.}
\end{table}

The inner-most radius was fixed at 100~mm.  Although it is
possible to construct telescopes with much smaller shells (see
table~\ref{tab:optics}), the net gain in collecting area is
minimal. For example, the telescope with F = 7.6~m would require
an additional 51 layers to populate the pupil annulus with inner
radius 50~mm and outer radius 100~mm, and the geometric area of
the telescope would only increase 4\%.

Next, we convolved the incident differential axion spectrum (shown
in figure~\ref{axion_flux}) with telescope response, assuming that
the telescope was coated with either nickel, gold or iridium.  We
assumed detector QE was constant as a function of energy and that
no buffer gas was present in the magnet bore.  In this scenario,
the nickel coating produced the highest number of focused photons
in the 0.1--10~keV band-pass, so we adopted nickel as the baseline
coating material.  Higher-Z materials, like Au or Ir, could become
the preferred coating material if the energy dependence of the
detector QE is considered or if system performance was optimized
for higher-mass axion searches.

After accounting for obscuration and realistic surface roughness
for a nickel coating, we computed a system efficiency
$\epsilon_{o}$ of between 28--37\% for the F = 5.7~m design and
37--49\% for the F = 7.6~m design.  The lower value assumes that
75\% of the x-rays properly reflected by the telescope falls
within a focus spot with a radius derived from
equation~\ref{eq:optic2} and listed in table~\ref{tab:optics2}.
The higher value assumes that 100\% of the reflected light falls
within the focus spot.  The actual efficiency of telescope will
fall somewhere within this range and depends on the manufacturing
technique.

This exercise demonstrates that the parameters assumed in
table~\ref{scenarios} could be achieved with a dedicated x-ray
optics fabrication effort for the NGAH.  An interesting
consideration is whether we can directly leverage any of the work
or infrastructure used to construct these telescopes.  For
example, eRosita will re-use the mandrels originally produced for
ABRIXAS \cite{pab+10}. This approach would only work for NGAH if
an exact duplicate would be appropriate.  This is because the
replication process does not allow for easy modification of the
master
 mandrels.  Shaping mandrels and other tooling are required
 to make either segmented glass or aluminum substrates,
 and this may be a more suitable hardware to exploit
 for the NGAH.  For example, the equipment used to
 make NuSTAR might be available for new
 projects, once all the flight hardware is complete.
 This tooling has more flexibility and could be
 more readily adapted to the specification of the NGAH optics.

\section{Detectors} \label{detectors}

\begin{figure}[ht!]
    \centering
     \includegraphics[width = .70\textwidth, angle = 0]{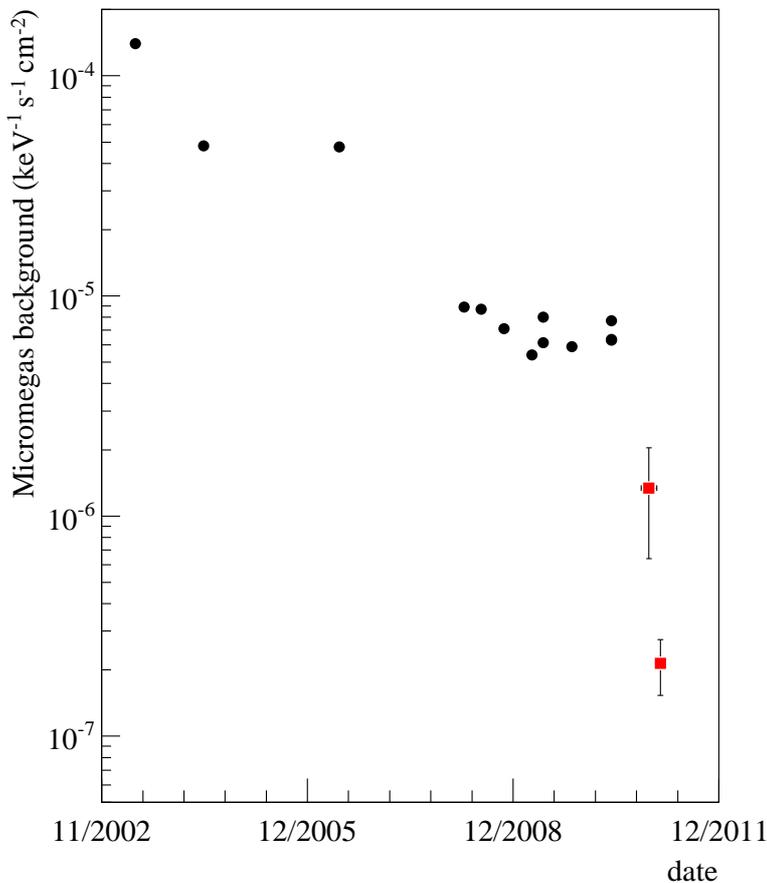}
      \caption{Background levels of Micromegas detectors over the years.
      Black points represents nominal values in CAST data taking campaigns.
      Squared red points correspond to data taken in special shielding conditions in the Canfranc underground laboratory.}
    \label{fig:mmback}
\end{figure}

There has been a continuous effort in CAST to improve the
background of the x-ray detectors. All CAST detectors, to some
extent, have adopted successful measures in this respect, like
shielding or low radioactivity materials
\cite{Abbon:2007ug,Kuster:2007ue,Autiero:2007uf}. The CAST
Micromegas detectors are however the most relevant example of
this, because of the reduction in background achieved and the
potential for further improvement. This is illustrated in figure~\ref{fig:mmback}, commented later on, where the background levels
achieved by the CAST Micromegas detectors along the experiment
lifetime are shown. We focus our following discussion on the
status of the development of these detectors and the prospects of
achieving levels of background down to $10^{-7}$ c/keV/cm$^2$/s as
anticipated in previous sections.

Micromegas are gaseous detectors (the usual gas mixture being Ar
with a fraction between 2\% to 5\% of isobutane) and as such they
need to be coupled to the vacuum bore via thin windows, keeping
the pressure difference but at the same time letting the x-rays
pass. The x-rays interact in the conversion volume that, in our
case, needs to be at least 3 cm thick in order to keep a good
detection efficiency. The ionization then drifts towards the
proper Micromegas readout \cite{Giomataris:1995fq,
Giomataris:2004aa} which consists of a metallic micromesh
suspended over a pixellised anode plane by means of insulator
pillars, defining an amplification gap in the range 25--150
$\mu$m. The drifting electrons go through the micromesh holes and
trigger an avalanche inside the gap, inducing detectable signals
both in the anode pixels and in the mesh. It is known
\cite{Giomataris:1998rc} that the way the amplification develops
in a Micromegas gap is such that its gain $G$ is less dependent on
geometrical factors (the gap size) or environmental ones (like the
temperature or pressure of the gas) than conventional multiwire
planes or other types of micropattern detectors based on charge
amplification. This fact allows in general for higher time
stability and spatial homogeneity in the response of Micromegas.
In addition, the amplification in the Micromegas gap has less
inherent statistical fluctuations than that of multiwire
proportional chambers (MWPCs), due to the faster transition from
the drift field to the amplification field provided by the
micromesh~\cite{Alkhazov:1970fx}.

\begin{figure}[t]
    \centering
     \includegraphics[width = .40\textwidth, angle = 0]{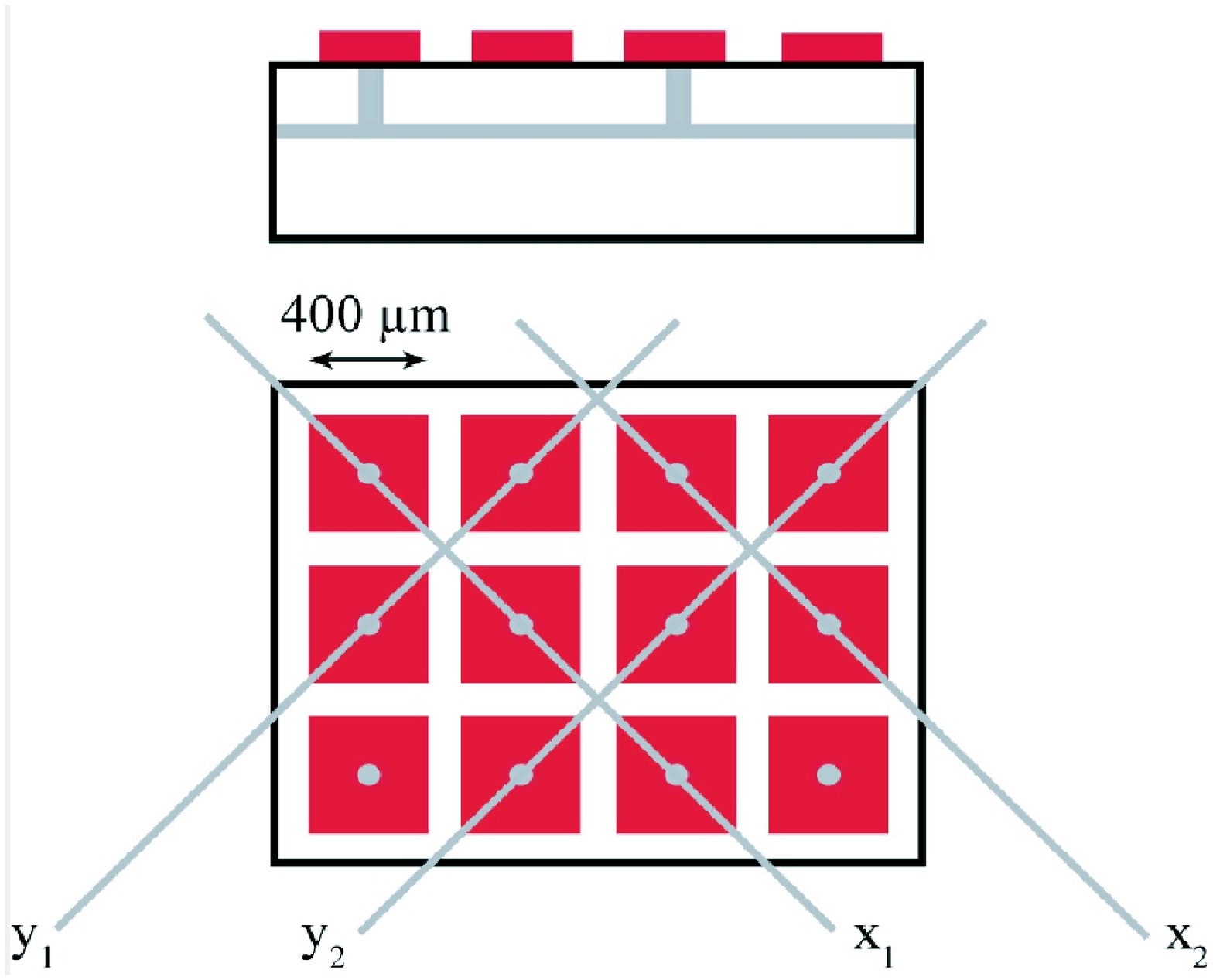}
      \includegraphics[width = .50\textwidth, angle = 0]{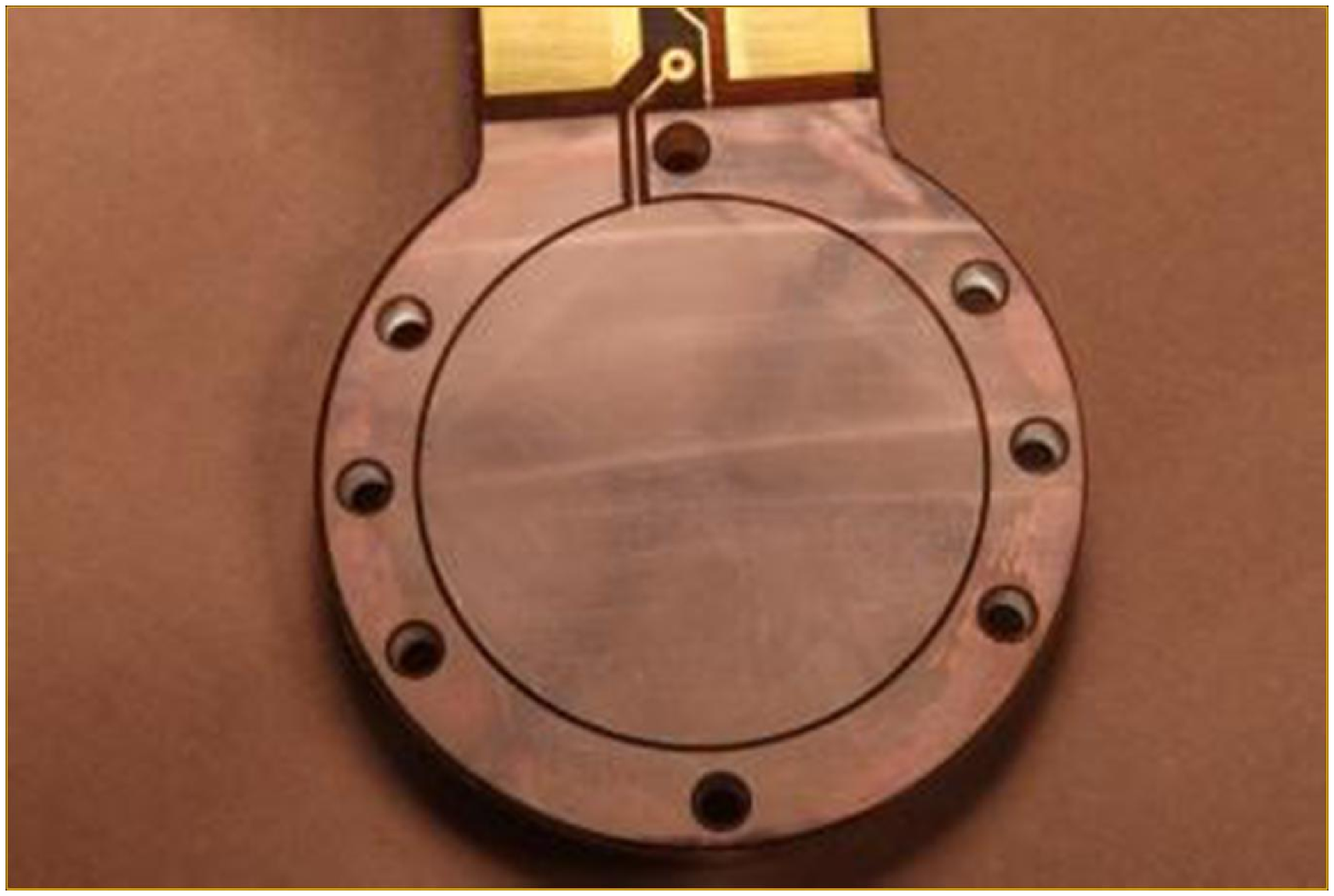}
      \caption{Left: Scheme of the 2-D readout plane of the Micromegas detectors used today in the CAST experiment. Right: Photo of the active area of a microbulk readout.}
    \label{fig:mm2d}
\end{figure}

The possibility of patterning the anode in pixels or strips of
very high granularity coupled with appropriate electronics allows
one to extract precious topological information of the event. This
fact, together with the rich topology offered by the gaseous
detection medium, has proven essential to design efficient
algorithms of signal identification and background substraction in
CAST~\cite{Galan:2010zz}. The 2-D readout pattern imprinted in the CAST Micromegas is
sketched in left panel of figure~\ref{fig:mm2d}, and is composed
by 400 $\mu$m side pixels linked in horizontal and vertical
strips. This granularity typically yields offline background
reductions of a around a factor 100 or more, depending on the
event energy.

The practical realization and operation of Micromegas detectors
have been extremely facilitated by the development of fabrication
processes which yield an all-in-one readout, in contrast to
``classical'' first generation Micromegas, for which the mesh was
mechanically mounted on top of the pixelised anode. Nowadays most
of the realizations of the Micromegas concept for applications in
particle, nuclear and astroparticle physics, follow the so-called
\emph{bulk}-Micromegas type of fabrication method or, more
recently, \emph{microbulk}-Micromegas (right panel of
figure~\ref{fig:mm2d}).

\begin{figure}[t]
    \centering
     \includegraphics[height = .40\textwidth, angle = 0]{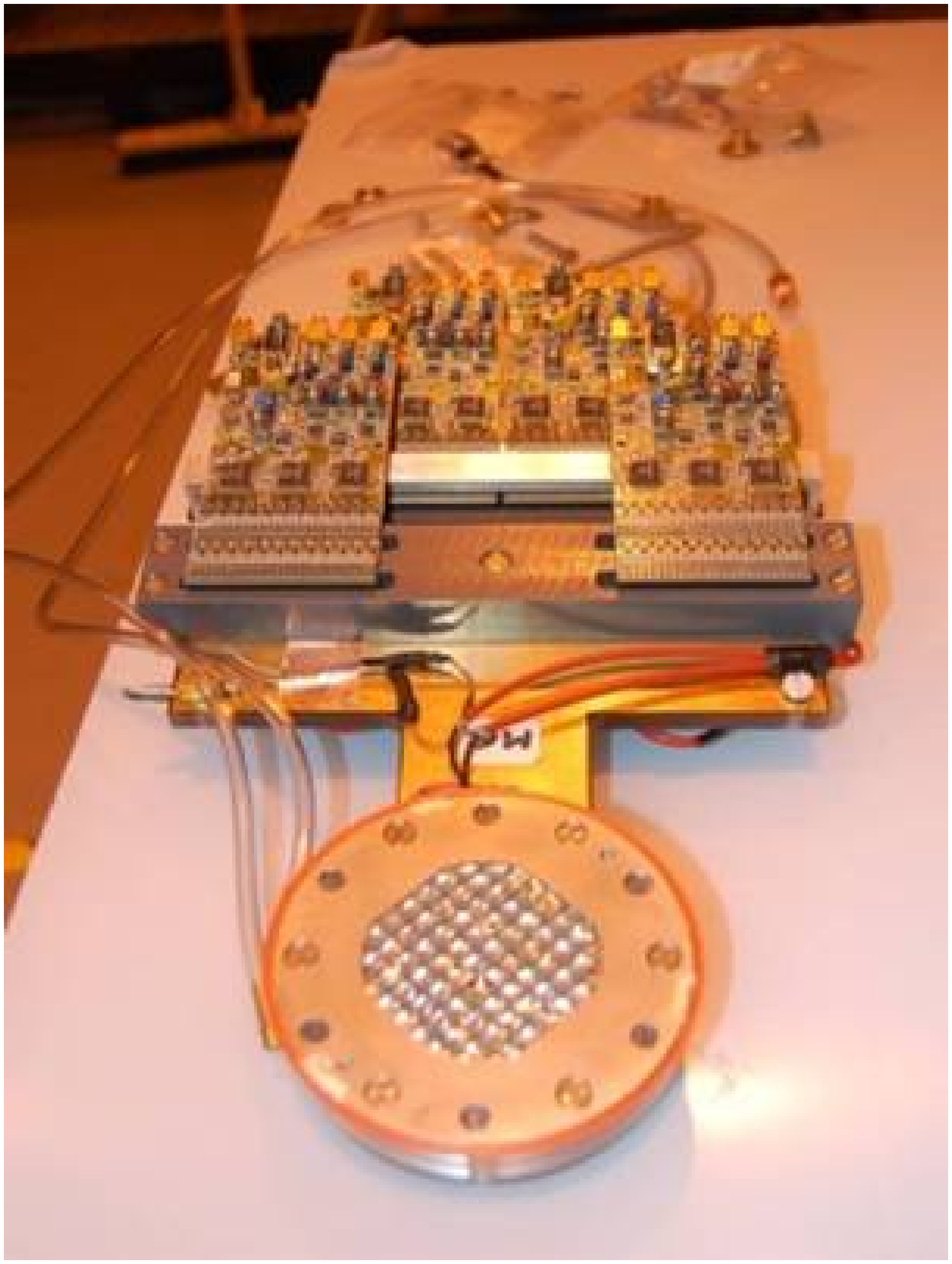}
      \includegraphics[height = .40\textwidth, angle = 0]{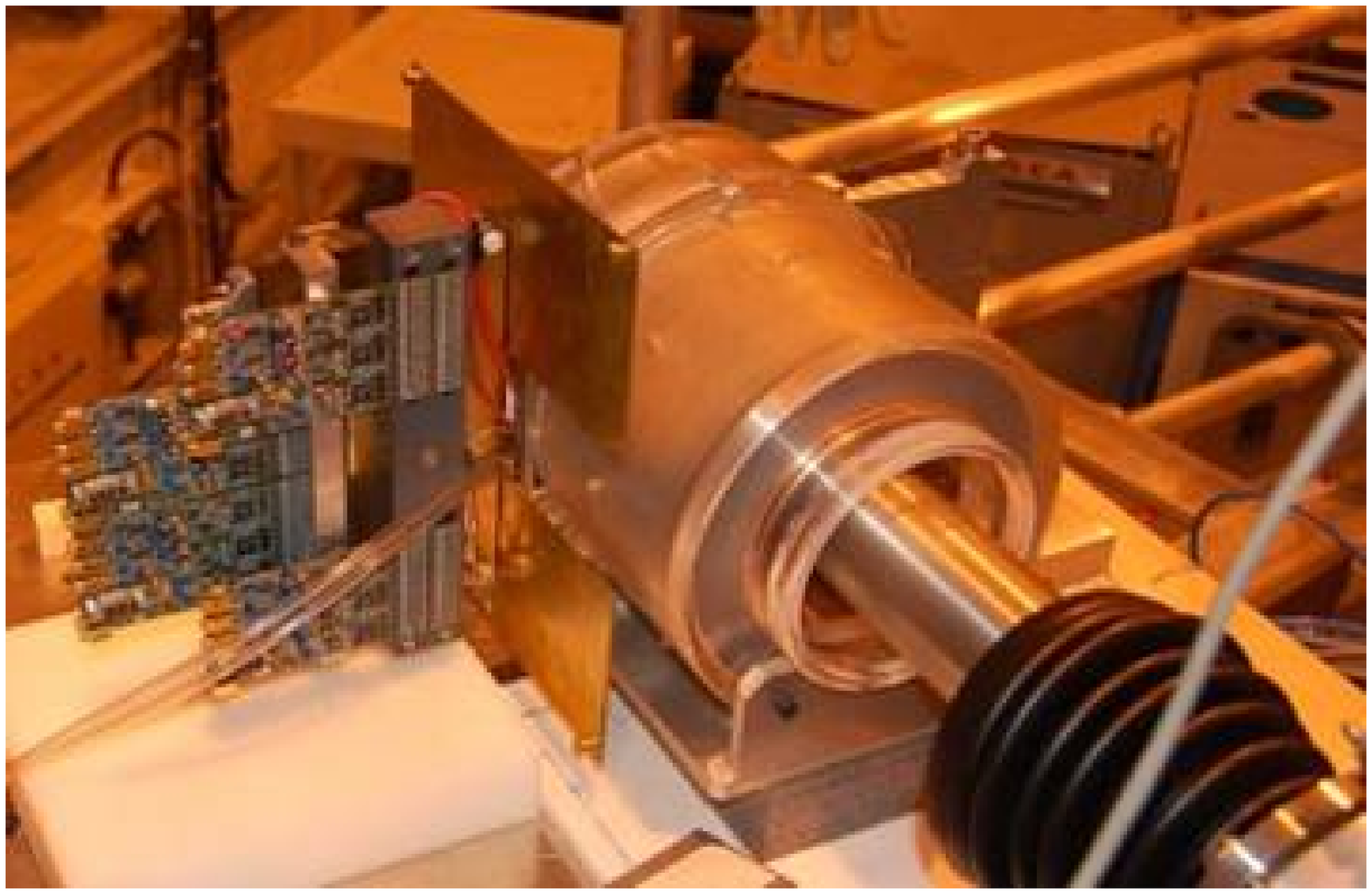}
      \caption{Left: Picture of a CAST Micromegas detector, with the x-ray window and readout electronics \cite{Abbon:2007ug}. Right: detector installed on the CAST magnet bore, surrounded by a lead cylinder, the inner part of the shielding.}
    \label{fig:mmpictures}
\end{figure}

While the bulk Micromegas \cite{Giomataris:2004aa} uses a photo
resistive film to integrate the mesh (usually a commercial woven
mesh) and anode, being already a mature and robust manufacturing
process, the microbulk Micromegas is a more recent development
\cite{Galan:2010zz,Andriamonje:2010zz}. It allows to provide, like
the bulk, all-in-one readouts but out of double-clad kapton foils.
The mesh is etched out of one of the copper layers of the foil,
and the Micromegas gap is created by removing part of the kapton
by means of appropriate chemical baths and photolithographic
techniques.
The mechanical
homogeneity of the gap and mesh geometry is superior, and in fact
these Micromegas have achieved the best energy resolutions among
MPGDs with charge amplification. Because of this, gain stability
of microbulk readouts is also superior. In addition,
the readout can be made extremely light and most of the raw
material is kapton and copper, two of the materials known to be
(or to achieve) the best levels of radiopurity \cite{ilias_db}.
Indeed, the first radiopurity study of Micromegas
\cite{Cebrian:2010ta} shows that current microbulk readouts
contain radioactivity levels at least as low as $57\pm 25$
$\mu$Bq/cm$^2$ for  $^{40}$K,  $26\pm 14$ $\mu$Bq/cm$^2$ for
$^{238}$U, $<13.9$ $\mu$Bq/cm$^2$ for $^{235}$U and $<9.3$
$\mu$Bq/cm$^2$ for $^{232}$Th. These levels are comparable to the
cleanest materials used in the detectors of the currently most
stringent low background experiments, e. g. in dark matter or
double beta decay experiments performed in underground
laboratories. Other materials composing the detector body are also
chosen to have low radioactivity levels (copper, plexiglass,
etc...).

Figure~\ref{fig:mmback} shows the background levels achieved by
the CAST Micromegas detectors along the experiment's lifetime.
Solid dots, representing the nominal levels achieved in CAST data
taking periods show a decrease in background by a factor 20 since
the start of the experiment. Last generation of Micromegas, made
with the microbulk fabrication technique, with radiopure
components and properly shielded (see
figure~\ref{fig:mmpictures}), present a background of 5--10$\times 10^{-6}$ \ckcs.


Micromegas are currently object of very active development to
understand their background limitations and eventually to improve
them. As part of this effort, several test benches with replica
detectors are in operation. Specially relevant is the one running
underground at the Canfranc Underground Laboratory (LSC). Data are
being taken testing diverse improvements in the shielding configuration. These data are being compared with detailed background simulation models under development, in order to understand ultimate background origins.
Some details of this effort have been presented in several specialized workshops \cite{Aune:2009zzb,Aune:2009zzc,Galan:2010zz,paco}.

Currently, as represented by the red points shown in
figure~\ref{fig:mmback}, background levels at the 1--5$\times
10^{-7}$~\ckcs range have been obtained, solely on the basis of
improvement on shielding thickness and coverage. This level
corresponds already to the one anticipated in the scenario NGAH-2 of
table \ref{scenarios}. Although work is still in progress,
preliminary considerations point to the fact that further
background reduction is possible beyond the one already obtained
via shielding improvements. The radiopurity of some materials
entering the detector components is still improvable (some glues,
connectors, feedthroughts and others). There is also margin to
improve at offline analysis, especially by going to better readout
electronics providing 3D topological information. Finally, the use
of an inner anticompton veto is also being studied. A quantitative
estimation of the expected improvement that these factors could
yield in background reduction is still pending, but a further
factor 3 or 10 (corresponding to scenarios 3 and 4 of
table~\ref{scenarios}) are certainly not unrealistic perspectives.
All these points will be explored in a dedicated prototyping R\&D
which is already ongoing.

\section{Conclusions}
\label{sec:conclusions}

We have shown that an enhanced axion helioscope, based on innovations
already introduced by CAST, could achieve a sensitivity of 1--1.5
orders of magnitude beyond current CAST limits. Specifically we have
reviewed the three key elements: the use of x-ray optics to increase
the signal-to-noise ratio, low background x-ray detectors, and a
toroidal magnet with a much larger geometric cross section. We have
found that there are realistic prospects to achieve the required
experimental parameters.

In terms of the axion-photon coupling constant $\gagamma$, this
instrument could approach the 10$^{-12}~{\rm GeV}^{-1}$ regime for
axion masses up to about 0.25~eV, covering completely unexplored
parameter space for general ALPs. At lower masses, in particular,
this region includes ALP parameters invoked repeatedly to explain
anomalies in light propagation over astronomical distances.

What is more, this experiment would cover a broad range of
realistic axion models that accompany the Peccei-Quinn solution of
the strong CP problem. If this instrument reaches its most
ambitious goals, the sensitivity would cover axion models with
masses down to the few meV range. It would supersede the SN~1987A
energy loss limits and could test the hypothesis that the cooling
of white dwarfs is enhanced by axion emission. We would explore
completely untested axion parameter space.

We therefore propose a next generation axion helioscope (NGAH),
following the design outlined above, as the next large-scale project
that the experimental axion community should embrace as a complement
to the ongoing  axion dark matter searches.

\acknowledgments

We thank our colleagues of the CAST collaboration. We acknowledge
support from the Spanish Ministry of Science and Innovation
(MICINN) under contract FPA2008-03456, as well as under the CPAN
project CSD2007-00042 from the Consolider-Ingenio2010 program of
the MICINN. Part of these grants are funded by the European
Regional Development Fund (ERDF/FEDER). We also acknowledge
support from the European Commission under the European Research
Council T-REX Starting Grant ERC-2009-StG-240054 of the IDEAS
program of the 7th EU Framework Program. Part of this work was
performed under the auspices of the U.S. Department of Energy by
Lawrence Livermore National Laboratory under Contract
DE-AC52-07NA27344 with support from the LDRD program through grant
10$-$SI$-$015. Partial support by the Deut\-sche
For\-schungs\-ge\-mein\-schaft (Germany) under grants TR-27 and
EXC-153, as well as by the MSES of Croatia, is also acknowledged.


\bibliographystyle{JHEP}
\bibliography{../../../../BibTeX/igorbib,pivovaroff,redondobib}

\end{document}